**Operando depth-resolved measurement of solvation entropy, interfacial transport and charge-transfer kinetics in lithium-ion batteries**

Divya Chalise[1,2], Sean Lubner[2,3], Sumanjeet Kaur[2], Venkat Srinivasan[4,*], Ravi S Prasher[1,2*]

[1] – Department of Mechanical Engineering, University of California, Berkeley, California, 94720, USA

[2] – Energy Technologies Area, Lawrence Berkeley National Lab, 1 Cyclotron Road, Berkeley, California, 94720, USA

[3] – Department of Mechanical Engineering, Boston University, Boston, Massachusetts, 02215, USA

[4] – Argonne National Laboratory, Lemont, Illinois, 60439, USA

[*] – Corresponding Authors: Venkat Srinivasan, vsrinivasan@anl.gov, Ravi Prasher: rsprasher@lbl.gov

**Abstract**

Understanding and improving the performance and longevity of lithium-ion batteries critically depends on insight into the dynamic processes occurring at buried electrode-electrolyte interfaces. However, direct, depth-resolved, and operando diagnosis of these interfaces remains a longstanding challenge due to their inaccessibility beneath bulk materials, the limitations of conventional surface- and bulk-sensitive characterization tools, and the difficulty of maintaining realistic cell environments during measurement. These challenges have made it nearly impossible to uniquely resolve important interfacial properties such as charge transfer resistance, SEI (solid electrolyte interphase) resistance, and solvation entropy at the individual electrode interfaces within a working cell, information that is essential for mechanistic insight and accelerated battery design. Here, we report the development of Modulated Electrothermal Sensing (METS), an operando technique that enables depth-resolved measurement of solvation entropy, interfacial transport resistance, charge-transfer resistance, and SEI resistance at individual electrode-electrolyte interfaces within practical lithium-ion batteries. By leveraging frequency-dependent, thermal-wave sensing and interface-specific modeling, METS uniquely attributes interfacial properties to specific electrodes, as validated by comparison with traditional electrochemical impedance spectroscopy (EIS). The ability to spatially and temporally resolve interfacial processes in real time provides new diagnostic capabilities crucial for mechanistic studies of battery degradation and for the rapid development of next-generation energy storage systems.

**Introduction**

Electrochemical energy storage devices, such as lithium-ion batteries, rely on complex ion and electron transfer processes that occur at buried electrode-electrolyte interfaces [1], [2], [3], [4]. These buried interfaces are critical to the mechanisms governing performance, aging, and failure, yet remain exceptionally difficult to probe directly, especially under realistic operating conditions [5]. The development of operando, depth-resolved diagnostics is fundamentally important because surface-sensitive tools are limited by minimal penetration depths, while bulk analysis techniques cannot localize interfacial phenomena, making it challenging to uniquely assign mechanistic signatures to particular electrode interfaces within working cells [5], [6], [7], [8], [9], [10] [11]. Operando depth-resolved measurements are especially valuable for battery research and development, as they enable real-time observation and quantification of how interfacial properties such as solvation entropy, ionic transport, and charge transfer kinetics evolve during cycling, SEI formation, and other transient processes, insights that are critical for guiding the design of longer-lasting, higher-performance batteries.

Various high resolution methods utilizing transmission electron microscopy [12], [13], [14], NMR [15], [16], [17], [18] and x-ray tomography [19], [20] have been developed for understanding these materials and interfaces in-situ however application of these techniques for operando measurement in a practical cell including multiple material and interfaces remains challenging. On the other hand, techniques such as electrochemical impedance spectroscopy (EIS) and voltage or capacity-based methods which are used for practical cell provide bulk average information but cannot directly provide multiple interface depth resolved information. The information from this technique is subject to interpretation based on presupposed assumptions that are necessary for avoiding non-unique fitting or for decreasing the number of fitting parameters, such as the assumption of symmetric (equal) resistance drop in chemically similar electrodes.

Unlike electrical (current/voltage) signatures of electrochemical processes, thermal signatures generated in wave-like form are critically damped [21] and decay exponentially with a characteristic length known as the thermal penetration depth ($\delta = \sqrt{(\alpha/\omega)}$, where $\alpha$ is the thermal diffusivity and $\omega$ is the frequency of the thermal wave). Thus, the spatial (depth) information is encoded in the frequency of the generated thermal signature. Consequently, the spatial origin of a particular thermal signature within a specific range of frequencies can be attributed to a specific layer or interface. This concept of the thermal penetration depth has been utilized to probe depth-resolved non-homogeneity in electrochemical processes by employing active thermal sensors that generate thermal waves and measure the temperature response of the underlying layers resulting from thermal property changes corresponding to physical changes in the cell [22], [23], [24]. For instance, non-uniformity in lithiation across porous insertion electrodes has been measured from the thermal conductivity change during lithiation/de-lithiation [22] and lithium interface morphology evolution in solid state lithium metal cells during lithium stripping/plating [23] has been measured from the thermal interface resistance. However, most electrochemical processes do not necessarily lead to measurable changes in thermal properties such as thermal conductivity, heat capacity or thermal interface resistance. Still, all electrochemical processes, from reversible or irreversible entropy change, generate heat [25], [26], [27], [28]. As illustrated in Figure 1 (a) and

Figure 1 (c), the process of the ion transport through the interfaces and electrolyte generates irreversible Ohmic (Joule) heat, the process of charge transfer at the interface generates irreversible non-Ohmic heat and the entropy of ion-solvation generates reversible entropic heat. If the thermal properties are known, it is possible to measure the magnitude and the spatial origin of the heat generation rate by measuring the oscillating temperature caused by the electrochemical processes [29] when the processes are excited at specific frequencies governed by the frequency of the current passed through the cell.

In this work we utilize various origins of heat generation to obtain relevant electrochemical information of an operational cell. Instead of sending thermal waves from an active sensor to probe physical changes, we generate thermal waves from the electrochemical processes themselves and measure the oscillating temperature rise with a passive sensor on the exterior of the cell as shown in Figure 1 (b) and (d). The spatial resolution in this measurement is obtained from the thermal penetration depth related to the frequency of the electrochemical-thermal signature and the process resolution (i.e. the identification of the electrochemical process leading to the thermal signature) is obtained from the harmonics and the current-heat generation rate relationship of the electrochemical-thermal signatures. The electrochemical process that lead to signal generation in lithium symmetric cells and in cells with porous electrodes are illustrated in Figure 1 (a) and 1 (c) respectively and their origin are discussed in the following discussion. A typical sensor and cell configuration for each cell are shown in Figure 1 (b) and 1(d) respectively. Because the thermal signatures are at multiple harmonics of the excitation current and are electrochemical in origin, while spatial resolution is achieved using frequency domain spectroscopy of the thermal signatures, we name this method Multi-harmonic Electro-Thermal Spectroscopy (METS). The temperature oscillations probed in METS are of the order of milli Kelvins and therefore require specialized thermometry [30], [31]. Additionally, the measured heat generation rates at the specific harmonics need to be related to specific electrochemical processes. Thus, in this work, we employ a highly sensitive thermometry utilizing phase-sensitive lock-in detection to isolate and measure the thermal signatures and subsequently develop the theoretical framework to interpret the thermal signatures as depth-resolved measurements of thermodynamic (entropic), kinetic (charge-transfer) and charge transport properties of the layers and interfaces in a relevant electrochemical system- namely lithium-ion batteries.

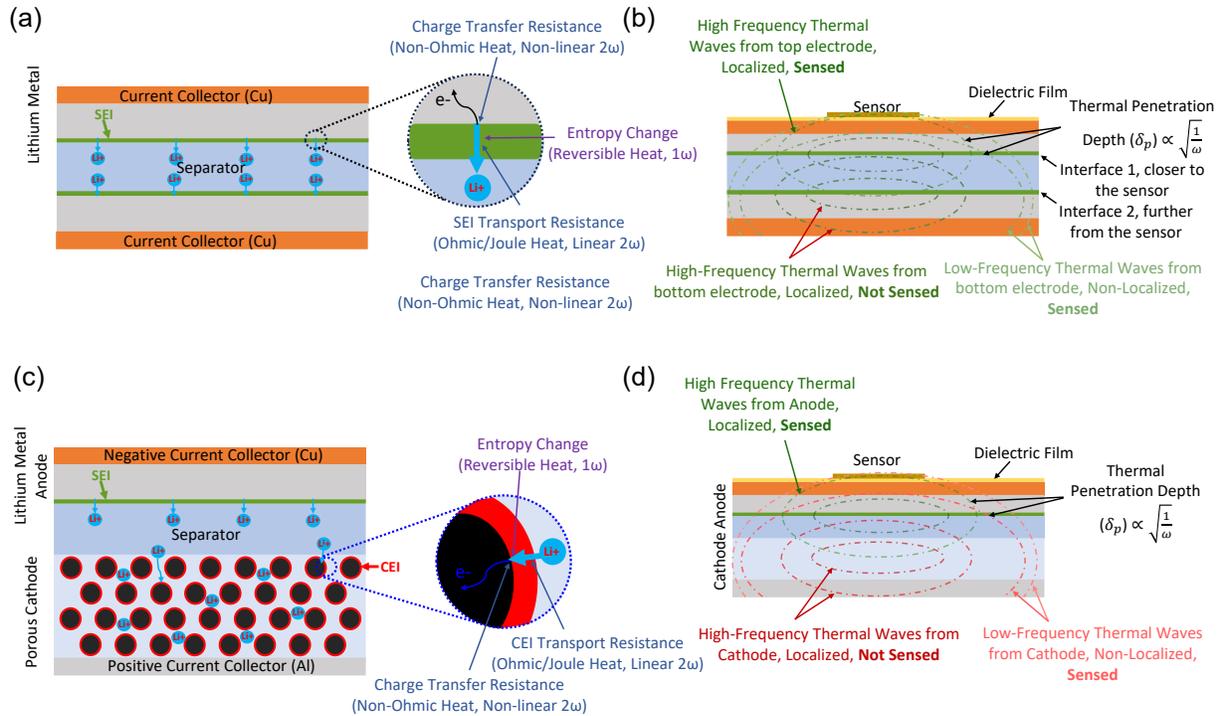

Figure 1. (a) Schematic of the symmetric lithium-ion cell with heat generated at the two electrode-electrolyte interfaces when an alternating current is passed through the cell. The expanded view illustrates the electrochemical processes leading to the heat generation at the electrode-electrolyte interface in the cell. The process of charge transfer at the interface generates irreversible heat which is non-Ohmic in nature. Additionally, the ion transport through the passivation layer (SEI) at the interface generates irreversible Ohmic (Joule) heat and the entropy of ion-solvation generates reversible entropic heat. (b) A simplified thermal schematic of the symmetric cell stack with the METS sensor used for the thermometry of the oscillating heat generation signatures. The interface closer to the sensor is identified as Interface 1 and the interface further away from the sensor is identified as Interface 2 and used consistently throughout the paper for all cells. The frequency of the current passed and consequently the frequency of the heat generated determines whether the generated is sensed by the sensor on the exterior of the cell. Heat generated at high frequencies are localized due to short thermal penetration depth. Therefore, at higher frequencies, only the heat generated at the interface close to the sensor is sensed. However, at lower frequencies, heat generated in both interfaces are sensed due to the long thermal penetration depths. (c) Schematic of a full cell with lithium metal anode and a porous cathode. In addition to the electrochemical processes at the anode-electrolyte interface, equivalent processes occur at the interface of the cathode particles and the electrolyte. However, unlike in the planar anode, the interfaces are present throughout the depth of the cathode because of its porous nature. (d) A simplified thermal schematic of the full cell stack with the METS sensor with the cathode treated as a uniform heat generation layer using homogenization of the porous layer (see text for details) from which the localized heat at higher frequencies are not sensed by the sensor while the heat generated at lower frequencies with longer penetration depths are sensed.

In a lithium-ion cell, when a lithium-ion moves from an electrode to the electrolyte, the solvation of the ion results in a significant entropy change [7]. This solvation entropy has been related to the practical aspects of battery design such as enhanced ionic conductivity and stability of lithium metal electrolyte interface [6], [32] and therefore, the measurement of the solvation entropy during battery operation enables monitoring degradation of the electrolyte and changes at the electrode-electrolyte interface. Previously, from calorimetry or temperature derivative of the open circuit potential [27], [33], it was only possible to measure the entropy change of the overall electrochemical reaction involving solvation of ions in one electrode and the desolvation in the other. The entropy changes of a half-cell reaction at a single electrode, i.e. the entropy of solvation/de-solvation at the electrode-electrolyte interface could not be measured. Recently, techniques developed by Wang et al. [7] and Cheng et al. [34] have made it possible to measure the entropy change of a single solvation/de-solvation reaction at the electrode-electrolyte interface. Nonetheless, the measurement developed by Wang et al. employs symmetric electrodes in a H-cell setup and cannot be performed in an operating cell. Similarly, the technique developed by Cheng et al. cannot resolve the difference in the solvation entropy at the two electrodes and is truly only applicable to symmetric cells. In this work, from the first harmonic ($1\omega$) thermal signature of electrode reactions performed at specific frequencies, we demonstrate a method capable of measuring and resolving the entropy of solvation at individual electrode-electrolyte interfaces in operational symmetric and non-symmetric cell. Details of mathematical formulism is given in SI.

Additionally, electrodes develop a passivation layer at the electrode-electrolyte interface commonly known as solid-electrolyte interphase (SEI) for anodes and cathode-electrolyte interphase (CEI) for cathodes. The presence of the passivation layer creates a transport resistance at the interface, and the growth of this passivation layer adversely affects the interfacial transport over time [35]. To understand how the passivation layer at each electrode evolves during battery operation, it is imperative to resolve the resistance at the two electrodes and track its evolution with time. While many non-operando studies [10], [36], [37] have investigated the SEI/CEI growth and evolution extensively, it is difficult to characterize the SEI/CEI evolution during the cell operation. Apart from the SEI/CEI transport resistance, there is an additional resistance at the interface related to the kinetics of the charge transfer. Separating the charge-transfer resistance from the interfacial transport resistance is challenging and has also been a topic of various studies[38], [39]. For electrodes with similar capacitance (pertaining to similar active surface area), EIS alone cannot resolve the interface resistance at one electrode versus the other and cannot separate the passivation layer transport resistance the from the charge transfer resistance, making the interpretation of the measured overall interfacial impedance ambiguous and subjective [40]. In this work, we utilize the second harmonic ($2\omega$) thermal signature to demonstrate operando measurement and resolution of the interfacial impedance into four different components, i.e. the charge transfer resistance and the interfacial transport resistance at each electrode. First, we use the thermal penetration depth of the second harmonic ($2\omega$) thermal signatures to spatially resolve the interfacial impedance in the two electrodes and then utilize the non-linearity in charge-transfer kinetics to separate the charge-transfer resistance from the interfacial transport resistance at individual electrodes, thereby unambiguously (and non-subjectively) resolving the overall interfacial impedance measured from EIS into the charge transfer resistance and interfacial transport resistance at the two electrodes. While doing this, we track the evolution of SEI resistance with battery ageing. More importantly demonstrate that the SEI resistance in two chemically similar (both lithium) electrodes can vary significantly based on the how the electrodes are prepared, highlighting the importance of the capability to resolve the interfacial resistance in the two electrodes in order to identify the defective electrode and obtain insights for improvement. Details of mathematical formulism is given in SI.

**First-harmonic (1ω) electro-thermal signature: Entropy of solvation at individual electrode-electrolyte interfaces**

The heat generation rate due to the entropy of solvation is reversible and equal to the product of the reaction current and the entropic coefficient related to the entropy change, i.e. $Q_{reversible} = -I_{rxn}T\left(\frac{dU}{dT}\right)$ where $I_{rxn}$ is the reaction current. $T$ is the absolute temperature and $\left(\frac{dU}{dT}\right)$ is the entropic coefficient [26], [27]. The entropic coefficient can be related to the entropy of solvation (i.e. entropy change of the half-cell reaction, $\Delta S_{rxn}$) by $\Delta S_{rxn} = nF\left(\frac{dU}{dT}\right)$. For a symmetric cell with both lithium electrodes and for a constant DC current passed, the entropy change in the two electrodes is equal and opposite to each other. Therefore, the overall entropic heating in the two electrodes is 0. When an AC current of frequency $\omega$ is passed through the cell, the entropic (reversible) heat generation rate, being proportional to the current, oscillates at the same frequency as the current, thereby creating a heating and temperature oscillation at frequency $\omega$ (or 1ω). The magnitudes of the 1ω heat at the symmetric electrodes are opposite to one another and therefore, at low frequencies (<<0.1 Hz) with long thermal penetration depth, the temperature oscillation due to opposite heating in the two electrodes cancel out each other. However, at higher frequencies, if the temperature sensor is placed on one end of the cell, closer to one electrode-electrolyte interface (referred to as Interface 1) than the other (referred to as Interface 2), the thermal penetration depth is short enough that the temperature oscillations created at the sensor by the entropic heating at Interface 1 is not cancelled out by the opposite entropic heating in Interface 2, as only the entropic heat from Interface 1 is sensed as shown in Figure 1 (a) and (c). It is therefore possible to measure the magnitude of 1ω entropic heat, and consequently, the entropic coefficient at each electrode from the frequency spectrum of the 1ω temperature measured by the METS sensor using the METS fitting algorithm presented in the SI.

For a 15mA alternating current passed through a symmetric cell with lithium metal foil electrodes and 1M LiPF$_6$ 1:1 EC:DEC (vol/vol) electrolyte, the measured 1ω temperature spectrum and the best-fit to the spectrum is presented in Figure 2(a). The in-phase and out-of-phase temperature rise are measured with reference to the phase of the alternating current passed through the cell. As presented in the discussion of the phase relationship between the current, heat generation rate and the measured temperature rise in Section 9 of the Supplementary Information (SI), the out-of-phase temperature rise corresponds to the sensible heat and is always positive for positive heating (and negative for negative heating). Therefore, at higher frequencies (>1 Hz), when the thermal penetration depth is short, the in-phase temperature rise is positive for a positive entropic heating at Interface 1. As the frequency decreases below 1 Hz, the thermal penetration depth becomes long enough, and the effect of the entropy change at Interface 2 is seen on the magnitude of the temperature oscillation at the sensor. Because the entropy change at the other electrode is equal and opposite, the out-of-phase temperature then starts getting cancelled as the frequency decreases, illustrated by the decreasing magnitude of the green circles in Figure 2(a) below 1Hz frequency. Theoretically, if the frequency decreases further, the out-of-phase temperature approaches zero as illustrated by the blue line in Figure 2 (a). However, because of practical consideration to avoid the possibility of dendrite formation in the lithium electrodes, experimental measurements were limited to 0.2 Hz and therefore are only presented till 0.2 Hz.

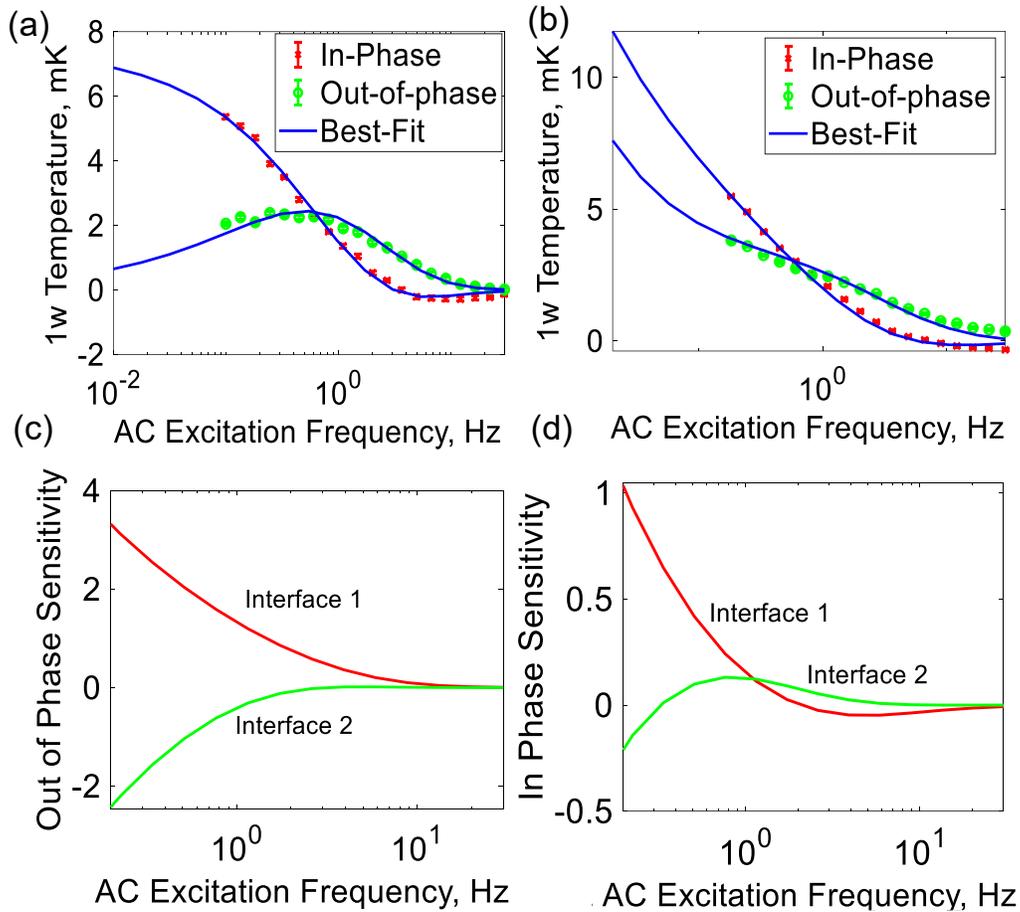

Figure 2. (a) 1ω temperature spectrum of the symmetric lithium cell plotted with the in-phase (red crosses) and out-of-phase (green circles) components of the temperature oscillation with respect to the alternating current passing through the cell. Going from higher frequency (30 Hz) to a lower frequency (1 Hz), the out-of-phase temperature, corresponding to the sensible heat, increases as the thermal penetration depth increases. However, at frequencies smaller than 1Hz, the opposite entropic heat at the other electrode is sensed, causing the temperature measurement to decrease and approaches zero at very low frequencies, as shown by the theoretical best-fit line (blue solid) The in-phase temperature, which is a function of thermal lag between the sensor and the heat generation site, does not decrease as the other effect of the other electrode in the in-phase temperature is minimal. (b) 1ω temperature spectrum of the NMC-lithium cell. Unlike in the case of the symmetric cell, the entropy change at the cathode and the anode are not equal, causing an imperfect cancellation and a residual out-of-phase temperature at low frequencies (<1 Hz), which keeps increasing with the increasing thermal penetration depth as the frequency decreases. Sensitivity plots for the out-of-phase (c) and in-phase (d) 1ω temperatures plotted as a function of frequency. As observed in the 1ω temperature plots, between 1 Hz and 30 Hz, frequency, the out-of-phase and in-phase 1ω temperature is only influenced by the entropic coefficient of the electrode-electrolyte interface closer to the sensor. At frequencies less than 1 Hz, the entropic coefficient of the other interface

(Interface 2) is also sensed in the out-of-phase measurement, while the effect of it on the in-phase measurement is negligible.

To explain the results further, we can define the sensitivity of the measurement to a measurement parameter ($p$) using the sensitivity term $S_p = \frac{dlnM}{dlnp} = \frac{p}{M}\frac{dM}{dp}$, where $M$ is the measured signal (in this case either in-phase or out-of-phase $1\omega$ temperature). The sensitivity can also be interpreted as the percentage change in the signal when the measurement parameter changes by 1%. This is helpful to illustrate the effect of the entropic coefficients of the two electrodes in the measured in-phase and out-of-phase $1\omega$ spectrum. From the $1\omega$ out-of-phase sensitivity plot (Figure 2(c)), above 1Hz, the out-of-phase $1\omega$ measurement is only sensitive to the entropic coefficient of Interface 1. However, at lower frequencies, the measurement is also sensitive to the entropic coefficient of Interface 2, where the entropy change is opposite, leading to a drop in the measured magnitude and consequently a negative sensitivity. Unlike the sign of the out-of-phase $1\omega$ temperature measurement, which only depends on the magnitude of the heating, the sign of the in-phase temperature measurement also depends on the thermal conduction lag between the heat generation site (in this case the two interfaces) and the sensor. From Feldman's solution [29], for a positive heating magnitude, if the depth probed is shorter than the thermal penetration depth ($\delta = \sqrt{(\alpha/\omega)}$), the sign of the in-phase $1\omega$ temperature rise is positive. However, if the depth probed is longer than or comparable to the thermal penetration depth, the out-of-phase temperature rise is negative. Accordingly, as seen in the in-phase sensitivity plot (Figure 2d) and the in-phase temperature measurement plot (Figure 2a), for Interface 1, at higher frequencies (>1Hz, short penetration depth), the in-phase temperature rise is negative while at lower frequencies (<1Hz, longer penetration depth), the in-phase temperature rise is positive. For interface 2, which is at a further distance from the sensor, the heating magnitude is opposite (negative). At the higher frequencies (>0.3Hz), the depth probed (i.e. the distance between the sensor and the interface) is longer than or comparable to the thermal penetration depth, the out-of-phase temperature rise is positive (for negative heating), which is evident in the sensitivity plot (Figure 2c). At frequencies lower than 0.3Hz, the sensitivity and therefore the temperature magnitude caused by the entropic heating at Interface 2 is negative. However, the overall sensitivity of the in-phase signal to the entropic coefficient of interface 2 is small and close to 0. Therefore, the in-phase signal in the overall spectrum and the out-of-phase signal at higher frequencies (>1Hz) can be uniquely fit with the entropic coefficient of Interface 1. Once the entropic coefficient of interface 1 is determined, the low frequency (<1Hz) out-of-phase signal can be used to determine the entropic coefficient of the second interface (Interface 2). For the symmetric cell, the two entropic coefficients must be the same. From the best-fit to the $1\omega$ spectrum, we determined the entropic coefficient for the lithium metal-electrolyte interface to be 1.2 ± 0.03 mV/K pertaining to the solvation entropy of 115.8 ± 3.4 J/molK. This value is very close to the values measured by Wang et al. (1.139 mV/K)[7] and Cahill et al. (1.04 mV/K for 1M LiPF$_6$ in EC-DMC)[34], which validates the accuracy of the measurement.

If the electrodes are not symmetric, the entropic coefficient at the two electrode-electrolyte interfaces are not the same. Therefore, the out-of-phase signal is not cancelled out perfectly. The high frequency in-phase and out-of-phase signals are still sensitive only to the entropic coefficient of the first interface while the low frequency in-phase and out-of-phase signals are sensitive to both. Additionally, the residual out-of-phase signal at low frequencies is proportional to the difference in the entropic coefficient at the two electrode electrolyte interfaces as the total heat generation rate is not perfectly cancelled. The $1\omega$ temperature spectrum for a lithium-ion cell with NMC 532 cathode at a state-of-charge (SOC) of 20%, lithium metal anode and 1M $LiPF_6$ 1:1 EC:DEC (vol/vol) electrolyte with the METS sensor places at the anode side is shown in Figure 2 (b). As the solvation entropy at the cathode-electrolyte interface and at the anode electrolyte interface are not equal, the out-of-phase temperature is not cancelled at lower frequencies (<1Hz), and theoretically keeps rising as the frequency increases. Further, from the best-fit to the METS spectrum at higher frequencies, we determine the entropic coefficient at the anode-electrolyte interface as $1.3 \pm 0.034$ mV/K ($\Delta S = 125.4 \pm 4.3$ J/molK and at the cathode electrolyte interface as $1.0 \pm 0.04$ mV/K ($\Delta S = 96.4 \pm 3.9$ J/molK). Note that the difference between the entropic coefficients at the two interfaces is $0.3 \pm 0.07$ mV/K, which is within the reported range (0.2-0.3 mV/K) of the entropic coefficient of NMC 532 lithium cells at 20% SOC [41], although this value could not be resolved into the entropic coefficients of the cathode and the anode in the previous measurement.

To show the interplay between the thermal penetration depth and opposing entropic coefficients further, we constructed a NMC523-graphite cell with sensors on both sides of the cell. For a particular SOC of 0.5 (OCV=3.5V), the theoretical entropic coefficient for a graphite anode is 1.05 mV/K [42] and for a NMC523 cathode cell is 1.28 mV/K [41]. The best-fit for the entropic coefficient for both the cathode-side sensor and anode-side sensor (Figure 3 (a) and (b) respectively) are obtained at $1.05 \pm 0.14$ mV/K for anode and $1.33 \pm 0.09$ mV/K for cathode, with the difference being 0.28 mV/K, which is close to difference between the cathode and anode values in the literature (0.23 mV/K). As the entropic coefficient of the cathode is larger than that for the anode, for the cathode side measurement, for low frequencies, the cathode side temperature rise dominates and continues to be positive as the frequency decreases. Complimentarily, for the anode side measurement, the opposite (cathode) side heating dominates, leading to a negative (opposite to the reference) temperature rise at low frequencies that correspond to the longer thermal penetration depths. For both sensors, the high frequency temperature rise is positive, as they only sense the heating from the adjacent electrode and cannot sense the cancellation from the opposite electrode.

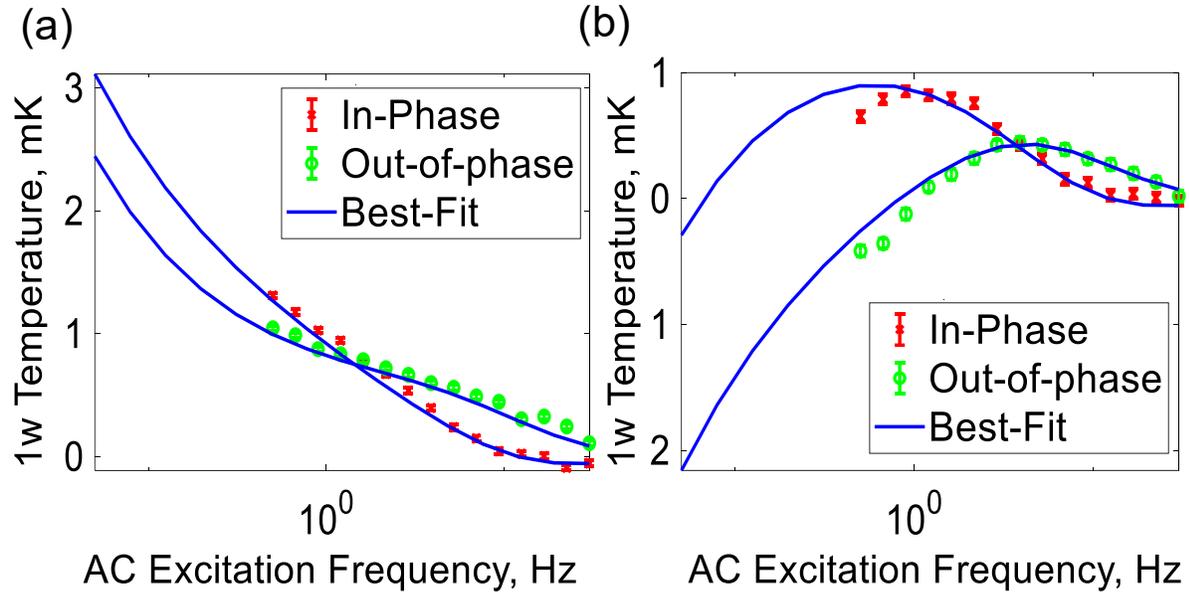

Figure 3. Thermal penetration depth and entropic coefficient interplay in an NMC523-graphite cell with dual-side sensors at 50% SOC with the measurement from the cathode side sensor (a) and from the anode side sensor (b). Experimental best-fit entropic coefficients are obtained at 1.05 ± 0.14 mV/K for anode and 1.33 ± 0.09 mV/K for cathode. Low-frequency measurements show cathode-side dominance (positive temperature rise at the cathode side sensor and negative temperature rise at the anode side senor) due to higher entropic coefficient at the cathode. High frequency measurements reflect localized heating at the adjacent electrodes resulting in positive temperature rise at both sensors.

**Second-harmonic (2ω) electro-thermal signature: Charge transfer resistance, SEI transport resistance and SEI growth**

The second harmonic thermal signature contains information about irreversible losses in the cell, specifically due to the transport resistance of the electrolyte, charge transfer resistances at the interfaces and the SEI/CEI transport resistance at the interfaces. Since the electrolyte transport resistance can be uniquely determined from the high-frequency intercept of the EIS bode-plot [40], we use METS 2ω spectra to determine the charge transfer resistance and transport resistance at the interfaces. As with the 1ω spectrum, we utilize the frequency dependence of the thermal penetration depth to resolve the processes at the two electrode-electrolyte interfaces. However, the heat generated due to charge transfer resistance and due to the SEI/CEI transport resistance at the same interface cannot be resolved spatially. To resolve this, we utilize the difference in the current-voltage relationship to separate the heat generation due to the transport resistance and the charge transfer resistance. The linearity in the current-voltage relationship (ohmic behavior) in transport resistance leads to the magnitude of the heat generation to scale with the square of the magnitude of the current while the non-linearity in the current-voltage relationship for charge transfer (described by the Butler-Volmer relationship [43]) leads to the heat generation magnitude scaling with the factor smaller than square of the current magnitude, thereby enabling unique fits of the charge transfer resistance and the transport resistance to the METS spectrum at different current magnitudes.

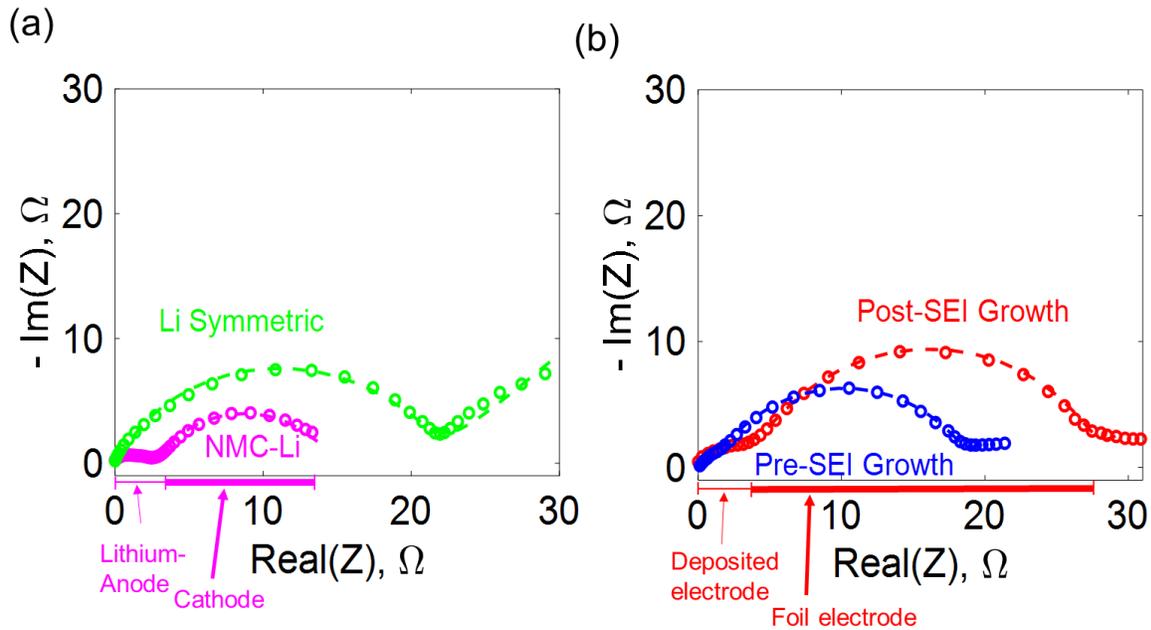

Figure 4. (a) EIS bode plots of the impedance of the symmetric lithium cell (green circles) and the NMC-Lithium cell (magenta circles). For the symmetric cell, with electrodes having similar capacitances, the overlapping semi-circles cannot be resolved and designated to a particular electrode-electrolyte interface, while for the NMC-Lithium cell, because of the dissimilar capacitances of the planar lithium anode and the porous NMC cathode, the larger semi-circle corresponding to lower frequencies can be attributed to the NMC cathode and the smaller semi-circle can be attributed to the lithium electrode. (d) EIS bode plots for the electrodeposited lithium-foil lithium cell with the measurement taken before the SEI growth (pre-SEI, blue circles) and after SEI growth (post-SEI, red circles). Although two overlapping semi-circles can be observed in both spectra, it is not possible to designate the semi-circles to a particular interface from the EIS measurement alone. However, from METS measurement, it is possible to unambiguously attribute the larger semi-circle to the foil electrode and the smaller semi-circle to the electrodeposited electrode.

For the symmetric cell, we measured the overall interface resistance to be $21.4 \pm 2.1$ $\Omega$ from EIS, shown in Figure 4 (a). Since the two electrodes are prepared the same way and assembled in the same cell, we expect the capacitance and the resistance of the two electrodes to be similar and therefore the semicircles observed in the EIS spectrum overlap each other. Since the interface resistances at each electrode cannot be clearly distinguished from EIS, we assume equal resistances at the two electrodes, which is half of the total resistance measured i.e. $10.7 \pm 1.1$ $\Omega$. From the best-fit to the METS $2\omega$ spectrum at 18mA, 20Ma and 22mA current, presented respectively in Figure 5 (a)-(c), the transport and charge-transfer resistance obtained for the interface closer to the sensor (Interface 1) were $9.02 \pm 0.84$ $\Omega$ and $0.5 \pm 0.88$ $\Omega$ respectively. Additionally, those values for the other interface (Interface 2) were $13.75 \pm 2.9$ $\Omega$ and $0.5 \pm 0.88$ $\Omega$ respectively. The total resistance measured for the two interfaces is $23.8$ $\Omega \pm 5.5$ $\Omega$ and is within the measurement uncertainty of the total interface resistance measured from EIS, which validates the accuracy of the METS $2\omega$ measurement.

For electrodes with dissimilar capacitances [44] such as the one with porous cathode and a planar lithium metal anode, it is possible to determine the interface resistances at the cathode-electrolyte interface and anode-electrolyte uniquely from EIS itself [45]. Therefore, a comparison of the interface resistances measured from EIS and METS can serve as an additional validation of the METS measurement. From EIS (presented in Figure 4(a)), for the same NMC 532 cathode-lithium metal anode cell presented in the 1ω analysis, we obtained the lithium-electrolyte interface resistance to be 2.26 ± 0.23 Ω and the cathode-electrolyte interface resistance to be 10.1 ± 1.0 Ω. From the 2ω METS spectrum at 18mA, 20Ma and 22mA current, presented respectively in Figure 5 (d)-(f), the transport and charge-transfer resistance obtained for the anode-electrolyte interface was 0.5 ± 0.11 Ω and 0.2 ± 0.11 Ω respectively. Similarly, those values for the cathode-electrolyte interface (Interface 2) were 12.2 ± 1.46 Ω and 0.2 ± 0.11 Ω respectively. The comparison of the EIS and 2ω METS measurements for the lithium symmetric cell and the NMC-lithium cell are presented in Table 1. It is evident that while the total interface resistance measured from EIS and METS are similar, METS can additionally resolve the resistance into four components, specifically the charge-transfer and the transport resistance at the two interfaces. More importantly, for all the measurements, the interface resistance is predominantly due to interface transport resistance, and the charge-transfer resistance is comparatively much smaller owing to relatively fast charge transfer kinetics.

Table 1. Summary of interfacial transport and charge-transfer resistances measured using METS and EIS for the symmetric lithium cell and the NMC-lithium cell

|  |  | **Electrode 1 (Lithium), Transport** | **Electrode 1 (Lithium), Charge-Transfer** | **Electrode 2 (NMC or lithium), Transport** | **Electrode 2 (NMC or lithium), Charge-Transfer** | **Total** |
|---|---|---|---|---|---|---|
| Lithium Symmetric | METS | 9.02 ± 0.84 Ω | 0.5 ± 0.88 Ω | 13.75 ± 2.9 Ω | 0.5 ± 0.88 Ω | **23.8 ± 5.5 Ω** |
|  | EIS | 10.7 ± 1.1 Ω | | 10.7 ±1.1 Ω | | **21.4 ± 2.14 Ω** |
| NMC-Lithium | METS | 0.5 ± 0.11 Ω | 0.2 ± 0.11 Ω | 12.2 ± 1.46 Ω | 0.2 ± 0.11 Ω | **13.1 ± 1.79Ω** |
|  | EIS | 2.26 ± 0.23 Ω | | 10.1 ± 1.0 Ω | | **12.36 ± 1.24 Ω** |

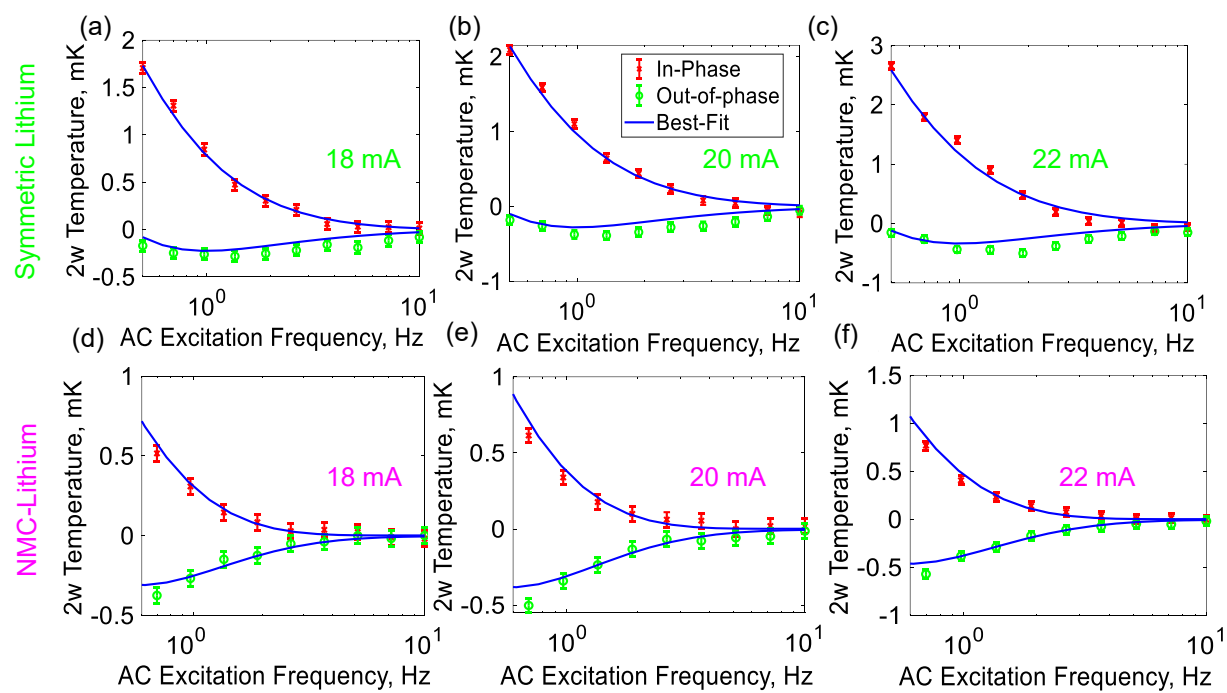

Figure 5. The 2ω temperature spectrum showing in-phase (red cross) and out-of-phase (green-circles) temperature rise as a function of the frequency of the current passed through the cell for current amplitudes of 18mA, 20mA and 22mA for the symmetric lithium cell (a-c, top) and the NMC-lithium cell (d-f, bottom). Unlike the 1ω measurement, 2ω measurements need to be performed at different current amplitudes to separate non-linear (non-Ohmic) charge transfer process with the linear (Ohmic) transport process.

When the electrode capacitances are similar, EIS cannot explicitly resolve the impedances in the two electrodes [45], [46]. This issue is more prominent when the two electrodes are chemically and morphologically similar so that there is no prior expectation of a particular electrode behaving in a certain way. Such is the case with an otherwise symmetric cell with a lithium foil electrode on one side and electrodeposited lithium electrode on the other. While having chemically similar electrodes (both lithium), the preparation of the two electrodes is different, leading to possibly different nature of the SEI resistance and the charge-transfer resistance at the two electrodes. To examine whether METS can unambiguously resolve the resistances at the two electrodes, we prepared a cell with a 100 μm lithium foil on one side and no electrode (bare current collector) on the other. We then electro-deposited 15μm lithium on the bare current collector side to create an electrodeposited lithium electrode on one side while having the foil electrode on the other. We first performed EIS measurement on the cell (shown in Figure 4 (b), pre-SEI growth measurement) and obtained the overall interface resistance to be 18.0 ± 1.8 Ω. The 2ω METS spectra measured at a specific current is shown in Figure 6 (a) and at three different current magnitudes are shown in the SI (Figure S12). The sensitivity for the in-phase and out-of-phase 2ω measurements are shown in Figure 6 (c) and 6 (d) respectively. Being performed at relatively low frequencies (<10 Hz), the measurement is not sensitive to the capacitance of the electrodes. Additionally, because of small electrolyte resistance and charge transfer resistance, the measurement sensitivities to these quantities are small compared to the interface transport resistance. As illustrated by the in-phase sensitivity plot (Figure 6(c)), and described in Section 9 of the supplementary information (SI), the in-phase temperature rise at the sensor, pertaining to sensible heating, is mostly sensitive to the transport resistance at the interface closer the sensor (Interface 1, electrodeposited electrode-electrolyte interface) at higher frequencies (>2Hz) corresponding to shorter thermal penetration depths and to the transport resistance at the interface further away from the sensor (Interface 2, foil-lithium electrode-electrolyte interface) at lower frequencies (<2Hz) corresponding to longer thermal penetration depths. The out-of-phase temperature rise at the sensor, related to the thermal conduction lag and represented by the green circles in Figure 6 (a) is negative and illustrates a significant thermal lag between the sensor and the heat source, indicating that the majority of the heat generation is at the interface away from the sensor, at Interface 2. Note that the phase relationship between the current and the temperature rise is reversed in the $2\omega$ measurements compared to the $1\omega$ measurements as the functional form of $2\omega$ heat 'cosine' while that of $1\omega$ heat is 'sine', as discussed in the supplementary information (SI). The best-fit to the $2\omega$ spectrum is achieved when the transport and the charge-transfer resistance at the electrodeposited electrode-electrolyte interface are $1.35 \pm 0.15$ $\Omega$ and $0.5 \pm 0.15$ $\Omega$ respectively and that at the foil electrode-electrolyte interface are $15.75 \pm 1.96$ $\Omega$ and $0.5 \pm 0.15$ $\Omega$ respectively. The sum of these resistances is equal to $18.1 \pm 2.41$ $\Omega$, which is close to the overall interface resistance measured from EIS (18 $\Omega$). Additionally, the charge transfer resistance in both interfaces is small, similar to the case with the symmetric cell and the NMC-lithium cell discussed earlier. More importantly, because of the spatial resolution, METS unambiguously shows that the interface transport resistance at the electro-deposited electrode is much smaller compared to that at the foil-electrode, most likely because of the pre-presence of surface impurities in the foil electrode, which is also illustrated by the EIS measurements on foil-foil and electrodeposited-electrodeposited symmetric cells presented in the SI (Figure S14).

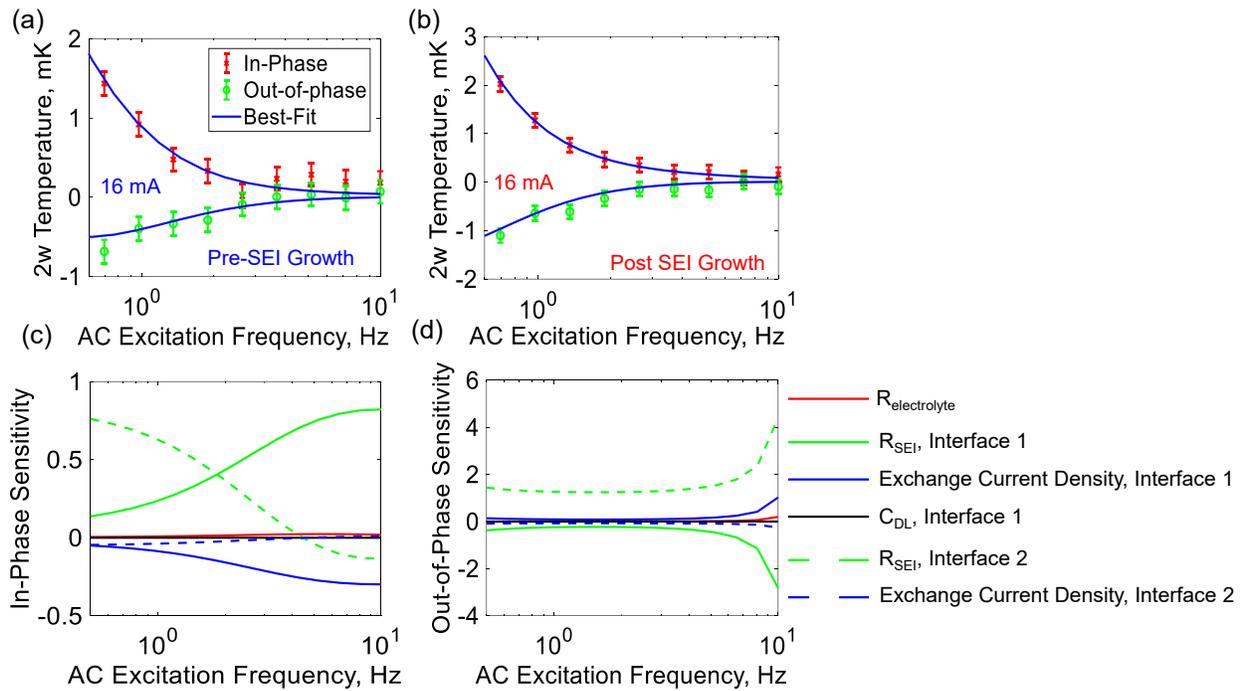

Figure 6. The 2ω temperature spectrum showing in-phase (red cross) and out-of-phase (green-circles) temperature rise as a function of the frequency of the current passed through the cell for and the same current amplitude of 16mA for the cell with one electrodeposited lithium electrode and one foil lithium electrode before SEI growth (a) and after SEI growth (b). The negative out-of-phase signal implies that the majority of the signal is generated at the interface away from the sensor, indicating that the resistance at the interface closer to the sensor is much smaller than the resistance at the interface away from the sensor. Additionally, the magnitude of the temperature for the same current magnitudes is seen to increase in (b) (post-SEI growth) when compared with (a) (pre-SEI growth) indicating the increase in Ohmic heat due to SEI growth. In-phase (c) and out-of-phase (d) sensitivity plots for the 2ω temperature measurements on the electrodeposited-foil lithium cell. Both in-phase and out-of-phase measurements are not sensitive to the electrode double layer capacitance and the electrolyte resistance in the frequencies of interest. Because of the fast-charge transfer kinetics (small charge-transfer resistance), both in-phase and out-of-phase temperature measurements are also not very sensitive to the exchange current density (conversely the charge-transfer resistance) at both interfaces. The in-phase temperature rise is mostly sensitive to the SEI resistance of the electrodeposited lithium-electrolyte interface (Interface 1) at higher frequencies (>2 Hz) corresponding to shorter thermal penetration depths and to the foil lithium-electrolyte interface (Interface 2) at lower frequencies (<2 Hz) corresponding to longer thermal penetration depths. Because of a larger magnitude, the SEI resistance of Interface 2 is much more sensitive in the out-of-phase temperature measurement than the SEI resistance of Interface 1, enabling unique fit to the SEI resistance in the out-of-phase temperature plot.

Finally, to examine the ability of METS to perform dynamic measurements in an operational cell, we created conditions for SEI growth [35] in the same cell with an electrodeposited lithium electrode and a foil-lithium electrodes by cycling 15μm equivalent lithium for 5 times at 40˚C. The EIS spectrum of the cell after the SEI growth (post-SEI growth) is presented as red circles in Figure 4 (b). The overall interface impedance from EIS is $26.9 \pm 2.7$ Ω, with two overlapping semi-circles with resistances 4.6 ± 0.46 Ω and 22.3 ± 2.23 Ω as shown in Figure 5 (b). The $2\omega$ METS spectra for the same cell for 16 mA current magnitudes is presented in Figure 6 (b) and that for three different current magnitudes are presented in the SI. Following a similar analysis as with the 'pre-SEI growth' case, the $2\omega$ METS spectra was fitted to obtain the transport and the charge-transfer resistance at the electrodeposited electrode-electrolyte interface to be 3.26 ± 0.22 Ω and 0.5 ± 0.38 Ω respectively and at the foil electrode-electrolyte interface to be 32.6 ± 5.3 Ω and 0.5 ± 0.38 Ω respectively. The comparison of the EIS and $2\omega$ METS measurements for the cell pre-SEI growth and post-SEI growth are presented in Table 3. The two semi-circles in the EIS spectrum (Figure 4 (b)) with resistances 4.6 Ω and 22.3 Ω, which could not be attributed to specific electrodes or process, can now be attributed to the transport resistances of the electrodeposited electrode and the foil electrode from the METS measurement, which highlights the importance of the spatial resolution enabled by thermal wave-based measurement.

Table 2. Summary of interfacial transport and charge-transfer resistances measured using METS and EIS for the cell with one electrodeposited and one foil lithium electrodes after SEI growth.

|  |  | Electrodeposited Electrode, Transport | Electrodeposited Electrode, Charge-Transfer | Foil Electrode, Transport | Foil Electrode, Charge-Transfer | Total |
| --- | --- | --- | --- | --- | --- | --- |
| Pre-SEI growth | METS | 1.35 ± 0.15 Ω | 0.5 ± 0.15 Ω | 15.75 Ω | 0.5 ± 0.15 Ω | **18.1 ± 2.41 Ω** |
|  | EIS | 1.9 ±0.19 Ω or 16.1 ± 1.61 Ω | | 1.9 ±0.19 Ω or 16.1 ± 1.61 Ω | | **18.0 ± 1.18Ω** |
| Post-SEI growth | METS | 3.26 ± 0.22 Ω | 0.5 ± 0.38 Ω | 32.6 ± 5.3 Ω | 0.5 ± 0.38 Ω | **35.8 ± 6.28 Ω** |
|  | EIS | 4.6 ± 0.46 Ω or 22.3 ± 0.22Ω | | 4.6 ± 0.46 Ω or 22.3 ± 0.22Ω | | **26.9 ± 2.7 Ω** |

**Discussion and Conclusions**

Unambiguous depth resolved measurement of electrochemical properties in an electrochemical cell is challenging and has not been demonstrated in operating cells. In this work, we have shown how the generation and measurement of thermal waves pertaining to electrochemical processes can enable the operando depth resolved measurement of the electrochemical properties such as interfacial charge transport, charge transfer and solvation entropy. The frequency spectrum of the temperature oscillation due to heat generation at the same frequency ($1\omega$) of the current passed through a cell can be used to measure and spatially resolve the solvation entropy at the two electrodes. Similarly, the frequency spectrum of temperature oscillation at the second harmonics ($2\omega$) of the current passed through the cell, taken at different current magnitudes, can be used to resolve the charge transfer resistance and the interface transport resistance at the two interfaces in the cell. The measurements carried out are minimally invasive and require simple instrumentation, enabling lab scale measurements in operational cells. As with the case of EIS, the instrumentation required for the measurement as presented in the SI is specialized nonetheless, as it requires isolation of low-amplitude signal from a large background noise, necessitating the use of frequency-based filtering either with a lock-in amplifier or using advanced numerical schemes. Therefore, we anticipate the use of this method for onboard measurements in an electric vehicle to be challenging without significant improvement in the cost and complexity of the instrumentation.

Finally, the measurements presented in this work are carried out at relatively high frequencies and ignore the mass transport related thermal effects. The method can also be extended to include the thermal signatures of mass transport at lower frequencies, opening the possibility of spatially resolved measurements of mass transport properties. Even though the results presented are only for lithium-ion cells, the method is generalizable to all electrochemical systems and can therefore be implemented in any electrochemical system where spatial resolution is important.

**Experimental Methods**

*Sensor Fabrication*

1in × 1in sections of thermally conductive Kapton® (McMaster) films (25μm thickness) were cut with protruding ends and 500nm copper was deposited on one side of the film to act as the current collector. On the other side, 4-point probe sensors (shown in Figure S7 (a) in the SI) were deposited as resistance thermometry (RTD) sensors. The sensors consisted of with a metallic line (150μm wide and 3mm long, or 300μm wide and 6mm long) and 4 attachment pads (2 each for passing current and measuring the voltage) and were deposited via subsequent e-beam evaporation of 10nm chromium and 100nm platinum through a laser-cut shadow mask. Electrical connections were made to the sensor pads by attaching 50μm diameter insulated copper wires using silver epoxy (EPO-TEK® H20E). The same sensor was used for 3ω thermal properties measurement and for METS 1ω and 2ω temperature measurements.

*Cell assembly*

Symmetric cells (Figure S7 (b)) were made by sandwiching 1in × 1in lithium foil electrodes (MSE Supplies) between 10μm copper current collectors deposited on Kapton® films with 25μm thick Celgard® 2400 separators in between. One of the dielectric films had the METS/3ω sensor deposited and wired. 2-3 mm thick Styrofoam sheet was attached on the sensor side of the cell to work as thermal insulation [47], [48] and a 2-3 mm thick Teflon plate was used on the other side to work as scaffolding. The cell was then sealed in a pouch cell configuration [47], [48] after adding the electrolyte (1M LiPF$_6$ in 1:1 EC:DEC, Sigma). NMC-Lithium full cell was made by using NMC-532 Cathode (MTI Corporation) with 60μm thick electrode and 15μm aluminum current collector. The sensor was placed on the lithium (anode) side. Cells with one electrodeposited lithium electrode and one foil electrode cells were made by electrodepositing 15μm lithium on the sensor side current collector from the lithium foil used on the other side. The thickness of the copper current collector on both sides was 0.5μm and Styrofoam was used on both sides of the cell instead of Teflon on one side. The stack configuration and the pouch cell assembly are illustrated in Figure S7 (c) in the Supplementary Information (SI). NMC-graphite full cell with sensors on both anode and the cathode side were made with NMC 532 Cathode (MTI) and 60μm thick graphite anode with 11μm copper current collector (MTI Corporation). Teflon plate was used on both sides as scaffolding/insulation. Multilayer NMC-graphite cell was made by folding a 2 in × 1in NMC-graphite cell in half to make a 1 in × 1in Cathode-Separator-Anode-Anode-Separator-Cathode multi-stack with the sensor inserted in the middle. The NMC-Lithium and NMC-graphite cells were subject to three constant current formation cycles with cutoff-voltages between 4.3V and 3.0V. The charge/discharge current for NMC-lithium, single layer NMC-graphite and multilayer NMC-graphite cells were 1.5mA, 1.1 mA and 2.4 mA respectively and the capacities achieved after formation were 12.5mAh, 3.5mAh and 9.5mAh respectively.

*High-precision thermometry instrumentation*

An in-house instrumentation was developed for frequency dependent (lock-in based) temperature measurements. The details of the instrumentation are presented in the SI. A constant DC current ($I_{DC}$) was passed through the sensor and half-bridge circuit with a matching resistor was implemented to cancel the dominant DC voltage across the sensor. The METS signal is generated by passing an alternating current at a frequency ω through the cell, which causes temperature oscillations at the sensor at frequencies 1ω and 2ω. The measured 1ω and 2ω voltages correspond to the sensor resistance $R_{1\omega}$ and $R_{2\omega}$ oscillating at 1ω and 2ω frequencies through the relation: $V_{1\omega} = I_{DC} R_{1\omega}$ and $V_{2\omega} = I_{DC} R_{2\omega}$. The measured 1ω and 2ω resistance is related to the oscillating temperature through the linearity in temperature dependence of sensor resistance i.e. $R_{1\omega} = \left(\frac{dR}{dT}\right) T_{1\omega}$ and $R_{2\omega} = \left(\frac{dR}{dT}\right) T_{2\omega}$, where $\left(\frac{dR}{dT}\right)$ is the linear temperature coefficient of resistance of the sensor. The schematic of the signal generation and instrumentation for the measurement is shown in the SI in Figure S8. Two Keithley 6221 current sources are used one as the AC source for the cell and the DC source for the sensor. The frequency of the AC source is referenced to a SR830 lock-in amplifier, which measures the voltage oscillations across the sensor. In all of our experiments, the typical noise in the voltage is within 100-200 nV. Considering the typical current through the sensor to be 10 mA and typical temperature coefficient of resistance of the sensor to be 0.15 Ω/K (varies slightly for each sensor), this voltage noise translates to a noise in temperature measurement of ~65-150μK.

*3ω measurements and thermal property characterization*

The same sensor and the setup used for METS signal acquisition is used for the measurement of the thermal properties using the 3ω-method by passing the alternating current through the sensor. The details of the 3ω method for the characterization of the thermal properties are presented in our earlier works [22], [23], [48] and explained in the SI. Thermal properties of each layers and interfaces used in the calculations are presented in Table S3 and Table S4 respectively and the 3ω fits for each cell are presented in Figure S9 in the SI.

*EIS measurements and galvanostatic cycling*

EIS measurements and galvanostatic cycling were carried out with Biologic MPG-2 Multichannel battery cycler. Potentiostatic EIS measurements were done between 200 kHz to 500 mHz with 2mV amplitude without any DC offset. In the case of electrodeposited cells, 15μm lithium was first electrodeposited on the sensor side current collector by passing 2mA current for 10 hours (20mAh total). To promote SEI growth in the cell with one electrodeposited electrode and one foil electrode, galvanostatic cycling was carried out at 40°C with 2 mA current with a voltage limitation of +-2.5V to cycle 15μm lithium 5 times. The formation and charging of the NMC-lithium cell was done using the same protocol described in our earlier work [22].

**Acknowledgements**

The authors acknowledge Dr. Yanbao Fu (LBL) and Dr. Vince Battaglia (LBL) for help with the cell assembly procedure, Dr. Kenneth Higa (LBL) for assistance with electrochemical measurements, Logan Vawter (LBL/UC Berkeley) for assistance with sensor fabrication, Dr. Drew Lilley (LBL/UC Berkeley) for assistance with the specific heat capacity measurement, Dr. Darshan Chalise (UIUC/Stanford) for the discussion regarding uncertainty analysis and Dr. David Cahill (UIUC) for the discussion regarding entropic/Peltier coefficients. The authors would also like to acknowledge the contribution of Joseph Schaadt (LBL/UC Berkeley) and Akshey Dhar (LBL/UC Berkeley) in the fabrication of earlier versions of the sensor. The MATLAB® script for the 3-omega analysis was developed by Dr. Yuqiang Zeng (LBL) for an earlier work [23]. This work was supported by the Assistant Secretary for Energy Efficiency and Renewable Energy, Vehicles Technology Office, of the U.S. Department of Energy under Contract No. DEAC02- 05CH11231.

**Declaration of Interests**

The authors declare that they have no known competing financial interests or personal relationships that could have appeared to influence the work reported in this article. Author DC is currently affiliated with Stanford University, and author RSP is currently affiliated with Bloom Energy. These affiliations are disclosed for transparency and do not constitute competing interests.

**Data Availability**

The METS, 3ω, EIS and thermal properties data used in this study are available upon request to the corresponding author.

**Code Availability**

The MATLAB® code used for the analysis of the data is available upon request to the corresponding author.

# Supplementary Information

## Depth-resolved measurement of solvation entropy, interfacial transport and charge-transfer kinetics of practical lithium-ion batteries

Divya Chalise[1,2], Sean Lubner[1,3], Sumanjeet Kaur[2], Venkat Srinivasan[4,*], Ravi S Prasher[1,2*]

[1] – Department of Mechanical Engineering, University of California, Berkeley, California, 94720, USA

[2] – Energy Technologies Area, Lawrence Berkeley National Lab, 1 Cyclotron Road, Berkeley, California, 94720, USA

[3] – Department of Mechanical Engineering, Boston University, Boston, Massachusetts, 02215, USA

[4] – Argonne National Laboratory, Lemont, Illinois, 60439, USA

[*] – Corresponding Authors: Venkat Srinivasan, vsrinivasan@anl.gov, Ravi Prasher: rsprasher@lbl.gov

**Table of Contents**

# 1. Overview of METS

When an alternating current at a frequency ($\omega$) is passed through an electrochemical cell, entropic, capacitive, charge transfer and charge transport processes result in heat generation either at the same frequency of the alternating current or at its harmonics. Entropic heat generation, being linearly proportional to the current, occurs at the same frequency as the alternating current ($\omega$). Other processes involve overpotentials (voltage drop) at the same frequency (for linear current-voltage relationship such as ohmic processes) or at multiples of the frequency (for non-linear current-voltage relationship such as charge-transfer processes) of the current, leading to irreversible heat, which is a product of the overpotential and the current, to be at the second harmonic ($2\omega$) and higher harmonics of the current. The heat generation rate because of these specific processes can be calculated from the electrochemical properties of interest using an appropriate circuit analysis, one of which is presented in the next section. These heat generation rates (at multiple harmonics) lead to a surface temperature rise at the same harmonics of the heat generation. However, because of the difference between the location of the origin of the heat signatures (i.e. the interfaces, electrodes and the electrolyte within the cell) and the location of the sensor (i.e. the outer surface of the cell), there is a time-lag between the generation and the sensing of the signature due to heat diffusion, which is manifested as a phase lag in frequency domain surface temperature rise and explained using an appropriate thermal analysis, such as Feldman's analysis [49] described in the next section. This phase lag carries the spatial information and can be used to attribute the thermal signatures to specific processes occurring at certain locations in the cell. The multi-harmonic temperature rise at the surface can be measured by resistance thermometry as voltage signal, which can be detected with a lock-in amplifier. This process is summarized in Figure S1 below. In the following sections, we will discuss the analysis and the experimental procedure in detail and explain how the obtained METS spectrum can be used to measure electrochemical properties of interest.

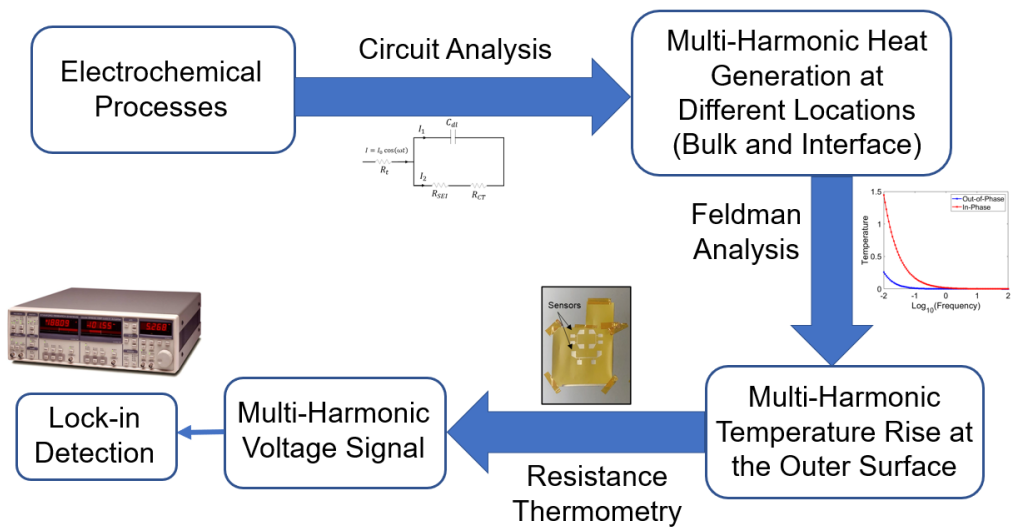

Figure S1. An overview of the METS method. When an alternating current at a specific frequency is passed through the cell, different electrochemical processes lead to thermal signatures at different harmonics of the excitation current. The multi-harmonic heat generation rates can be related to multi-harmonic surface temperature rise using appropriate thermal analysis such as Feldman's analysis [49] and, by resistance thermometry, can be measured as frequency domain voltage signal using a lock-in amplifier.

## 2. Frequency dependent heat generation from electrochemical processes

When a sinusoidal alternating current (AC) is passed through a cell, heat generation can occur due to reversible entropy change, irreversible losses, side reactions and mixing. In this analysis, we ignore the mass transport aspect of heat generation, the explanation of which is provided later in this document. Therefore, we ignore the heat of mixing. Additionally, we also do not consider heat generation due to side reactions as side reactions are not the predominant electrochemical reactions when an alternating current is passed through the cell. Therefore, we develop heat generation terms associated with entropy change, transport resistance and kinetic overpotential at interfaces and the transport resistance in the bulk electrolyte. In the following analysis, we demonstrate how the frequency dependent heat generation rate can be related to the electrochemical processes in a simple electrochemical model. However, the same method can be extended to more complex electrochemical models by modifying the equivalent electrochemical circuit and using an appropriate kinetic model for the electrochemical reaction.

*Example analysis on a model system*

Consider an electrochemical cell with planar electrodes separated by a separator/electrolyte. Then, assuming ideal capacitive behavior at the electrode double layer, the equivalent electrochemical circuit with one of the electrodes can be represented as the circuit represented in Figure S2.

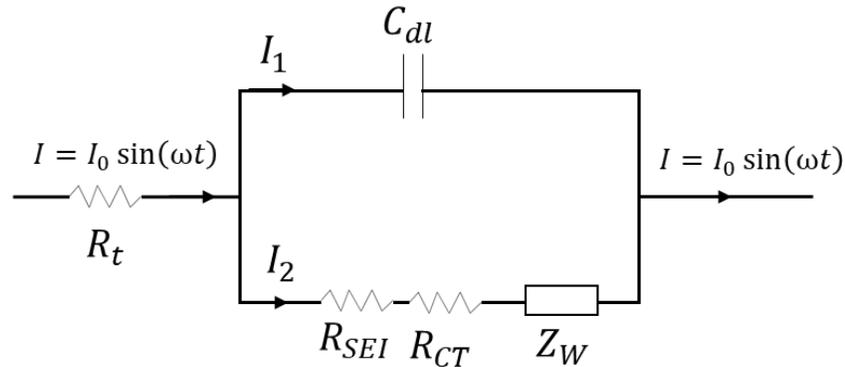

Figure S2. Equivalent electrochemical circuit for the electrolyte and a planar electrode with ideal double layer capacitance ($C_{dl}$), transport ($R_{SEI}$) and charge transfer resistance ($R_{CT}$) related to the reaction and a Warburg Impedance ($Z_W$) associated with concentration oscillation.

The resistance associated to charge transport in the electrolyte is denoted as $R_t$. The sinusoidal current passing through the cell $I = I_0 \sin(\omega t)$ can divided into two branches, capacitive, where the oscillating electric field can pass through the electric double layer capacitance ($C_{dl}$) and reactive, where the current can lead to a reaction at the electrode interface. The current through the capacitive branch is denoted as $I_1$ and through the reactive branch is denoted as $I_2$. The magnitude and phase of the branches will be calculated later. For a reaction to proceed at the interface, the ions move through a passivation layer called the solid electrolyte interphase (SEI), and the transport resistance associated with it is denoted by $R_{SEI}$. This resistance is ohmic in nature. Additionally, the overpotential associated with the charge transport kinetics leads to an additional resistance term denoted as $R_{CT}$. Additionally, there is another impedance term associated with the voltage oscillation as a result of concentration oscillation at the interface, and is known as the Warburg Impedance ($Z_w$). In our analysis, we ignore this Warburg Impedance based on the condition that we operate at a high enough frequency to avoid causing significant concentration oscillation. To determine the criteria to ignore the mass transport (concentration oscillation i.e. Warburg Impedance) effects, we use the principle of Sand's time ($\tau$) [50]:

$$\tau = \frac{\pi D_s}{4}\left(\frac{c_o nF}{s_i i}\right)^2 \tag{S1}$$

where, $D_s$ is the diffusivity of ions (lithium ion in the case of lithium-ion cells) in the electrolyte, $c_o$ is the nominal concentration of the electrolyte, $n$ is the number of electrons transferred, $F$ is the Faraday's constant, $i$ is the current density at the electrode and $s_i$ is the stoichiometry of the ion consumed in the electrochemical reaction.

The mass transport effect can be neglected if we operate within 10% of the Sand's time, i.e.[50]

$$t \leq 0.1\tau \tag{S2}$$

In frequency domain,

$$f \geq \frac{10}{\tau} \tag{S3}$$

In our experiments, the operating frequencies are chosen so that the inequality (S3) is always satisfied.

If so, the equivalent electrochemical circuit can be simplified as:

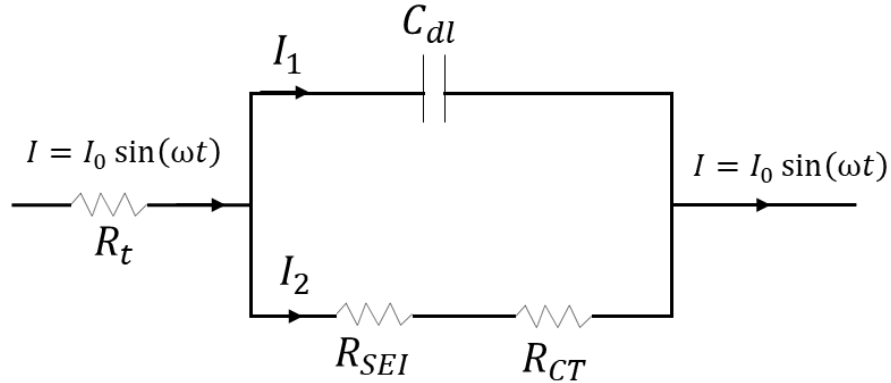

Figure S3. Simplified equivalent electrochemical circuit for the electrolyte and a planar electrode with ideal double layer capacitance ($C_{dl}$), transport ($R_{SEI}$) and charge transfer resistance ($R_{CT}$) without the Warburg element

To determine the magnitude and phase of the current distribution, we can apply a current divider formalism. Because of the non-linear current-voltage relationship associated with the charge transfer kinetics, the equivalent resistance associated with the charge transfer ($R_{CT}$) is current dependent. Thus, we cannot determine the current distribution directly and need to determine it iteratively. In our analysis, we use the Gauss-Seidel approach to determine the current distribution. Additionally, we assume Butler-Volmer kinetics to describe the current-voltage relationship for the charge transfer process. This is commonly used for lithium-ion charge transfer reactions [51], [52]. However, the same methodology can be applied when using other kinetic relations as well.

From the Butler-Volmer relationship, for the current $I_2$ passing through a cross-sectional area $A_s$, the overpotential $\eta_s$ associated with the charge transfer process is related to the current as:

$$i_2 = \frac{I_2}{A_s} = i_e \left( exp\left(\frac{\alpha_s F \eta_s}{RT}\right) - exp\left(-\frac{\alpha_s F \eta_s}{RT}\right) \right) \tag{S4}$$

where $R$ is the universal gas constant, $T$ is the absolute temperature, $\alpha_s$ is the symmetry-factor that relates the overpotential associated with the forward and the reverse reaction and is assumed to be 0.5 in our analysis, and $i_e$ is the exchange current density, which is a measure of the electrode kinetics.

Using this expression, the overpotential can be solved numerically. Since the overpotential and the current are non-linearly related, for a purely sinusoidal current at frequency $\omega$, the overpotential will have components at odd harmonics of $\omega$, i.e. at 1$\omega$, 3$\omega$, 5$\omega$ and so on as shown in Figure S4 and explained in the discussion following equations S22 and S23. Once the first harmonic of the overpotential is known, the equivalent charge transfer resistance can be calculated as:

$$R_{CT} = |\eta_{s,1\omega}|/|I_2| \tag{S5}$$

The impedance associated with the double layer capacitance can be written as:

$$Z_{dl} = -\frac{j}{\omega C_{dl}} \tag{S6}$$

Then, the current distribution at the interface can be determined from the current divider formula:

$$I_2 = \frac{\left(-\frac{j}{\omega C_{dl}} I_0\right)}{-\frac{j}{\omega C_{dl}} + |\eta_{s,1\omega}|/|I_2| + R_{SEI}} \tag{S7}$$

Since this expression (equation S7) contains $I_2$ on both sides, it can only be solved numerically. We choose the Gauss-Seidel approach to solve the current distribution iteratively.

Once solved, we obtain the magnitude ($I_{2,0}$) and phase ($\phi_2$) of the current $I_2$, i.e.

$$I_2 = I_{2,0} sin\,(\omega t + \phi_2) \tag{S8}$$

Similarly, $I_1$ can be solved as:

$$I_1 = I - I_2 = I_{1,0} sin(\omega t + \phi_1) \tag{S9}$$

In the typical frequency range for METS experiments, since $C_{dl}$ is small and $Z_1$ is large, $\phi_2$ is small and negative and $\phi_1$ is close to but smaller than $\pi/2$.

If we define: $\phi_{1c} = \pi/2 - \phi_1$ and $\phi_{2s} = -\phi_2$. Then,

$$I_2 = I_{2,0} sin\,(\omega t - \phi_{2s}) \tag{S10}$$

$$I_1 = I_{1,0} \cos(\omega t - \phi_{1c}) \tag{S11}$$

Since the voltage across the capacitive and the reactive branches has to be the same, it can be shown that $\phi_{2s} = \phi_{1c}$ and are small and positive in typical experiments.

Once the current distribution in each branch is determined, we can develop the heat generation terms associated with each electrochemical process. A similar analysis can be done to determine the current distribution at the other electrode.

*Heat generation due to entropy change*

At each electrode, during a charge transfer reaction, when an ion moves from the electrolyte into the electrode or vice versa, there is an entropy change associated with the charge transfer reaction. The heat absorbed or released because of this entropy change is proportional to the current. Newman et al. [53], [54] have shown that the heat generation due to the entropy change can be related to a term known as the entropic coefficient ($\frac{dU}{dT}$) which is related to the entropy change as:

$$\Delta S_{rxn} = nF \left(\frac{dU}{dT}\right) \tag{S12}$$

The reversible heat at the electrode with the reactive current $I_2$ is given by:

$$Q_{reversible} = -I_2 T \left(\frac{dU}{dT}\right) \tag{S13}$$

Since this heat is reversible and directly proportional to the current, the reversible heat can also be represented as a product of the current and the equivalent Peltier coefficient of the electrode ($\pi_{electrode}$) [55]:

$$Q_{reversible} = -I_2 \pi_{electrode} \tag{S14}$$

Since this reversible heat is directly proportional to the current at a frequency ω, the heat generated will also be at the same frequency ω. Additionally, since the current $I_2$ has a phase offset $\phi_{2c}$ compared to the reference frequency ω, this reversible heat can be divided into an in-phase (IP) and an out-of-phase (OP) component, i.e.

$$Q_{reversible,1\omega,IP} = -I_{2,0} \pi_{electrode} \cos(\phi_{2s}) \tag{S15}$$

$$Q_{reversible,1\omega,OP} = I_{2,0} \pi_{electrode} \sin(\phi_{2s}) \tag{S16}$$

*Heat generation due to transport resistance*

The resistance associated with transport is ohmic, i.e. the current voltage relationship for a transport process is linear. Then the heat generation rate for a current $I$ passing through a resistance $R$ can be calculated as $Q = I^2 R$.

In the electrolyte, the current $I = I_0 \sin(\omega t)$ passes through the resistance $R_t$. Thus, the heat generation rate (Watts, W) can be calculated as:

$$Q_{trasnport,electrolyte} = (I_0 \sin(\omega t))^2 R_t \tag{S17}$$

This heat will have components in DC (constant offset) and in the second harmonics (2ω). Also, the 2ω component of the heat is a cosine wave at twice the reference frequency ω, there is no in-phase (sine) component of the heat generation. The out-of-phase heat generation rate at the second harmonics (2ω) due to transport resistance at the electrolyte is calculated as:

$$Q_{transport,electrolyte,2\omega,OP} = -\frac{I_0^2 R_t}{2} \tag{S18}$$

At the electrode, the transport resistance associated with the ion transport through the SEI can be calculated as:

$$Q_{transport,SEI} = I_2^2 R_{SEI} \tag{S19}$$

This heat will also have components in DC and 2ω. Also, since the current $I_2$ has a phase offset $\phi_{2c}$ the second harmonic heat can be divided into an in-phase (IP) and an out-of-phase (OP) components, i.e.

$$Q_{transport,SEI,2\omega,IP} = -\frac{I_{2,0}^2 R_{SEI}}{2} \sin(2\phi_{2s}) \tag{S20}$$

$$Q_{transport,SEI,2\omega,OP} = -\frac{I_{2,0}^2 R_{SEI}}{2} \cos(2\phi_{2s}) \tag{S21}$$

*Heat generation associated with charge-transfer kinetics*

For a reaction current $I_2$, the heat generation rate associated with the reaction overpotential $\eta_s$ is given by:

$$Q_{kin} = I_2 \eta_s \tag{S22}$$

Since the Butler-Volmer relation is a non-linear current voltage relation, for a sinusoidal current $I_2$, the kinetic overpotential $\eta_s$ will have components at odd harmonics of ω, i.e. at 1ω, 3ω, 5ω and so on as shown below in Figure S4.. Thus, the heat generation rate, which is a product of the current and the overpotential will have components at DC, 2ω, 4ω and so on. This is illustrated in Figure S4, where for a sinusoidal current, the numerically solved overpotential is shown as a function of time in Figure S4 (a) and its frequency domain expansion (Fourier transform) is shown in Figure S4 (b). As seen, the overpotential is not purely sinusoidal as it contains other harmonics at odd multiples for the primary frequency (1ω). Similarly, the time-domain evolution of the heat generation rate from the product of the current and the overpotential is shown in Figure S4 (c) and its frequency-domain expansion is shown in Figure S4 (d). The heat generation rate has a DC offset (0 Hz component) as well as a dominant oscillating component at the second harmonic of the current i.e. at 2ω. It also has higher harmonic components at even multiples of the current frequency, but the normalized magnitudes of these components are small and therefore difficult to measure. Therefore, we restrict our measurement to the second harmonic.

To isolate the second harmonic, we can use the coefficient of Fourier expansion of the overall heat generation rate $Q_{kin}$, i.e.

$$Q_{kin,\,2\omega} = \frac{1}{\pi} \int_0^{2\pi} \cos(2\zeta)\, Q_{kin}(\zeta) d\zeta \tag{S23}$$

The phase of this heat generation with reference to the AC frequency ω will be $2\phi_{2c}$ as the phase of the reaction current is $\phi_{2c}$ and so is the phase of the first harmonic of the reaction overpotential.

Then, the in-phase and out-of-phase components of the heat generation rate associated with the reaction kinetics can be written as:

$$Q_{kin,2\omega,IP} = \left(\frac{1}{\pi}\int_0^{2\pi} \cos(2\zeta)\, Q_{kin}(\zeta)d\zeta\right)\sin(2\phi_{2s}) \tag{S24}$$

$$Q_{kin,2\omega,OP} = \left(\frac{1}{\pi}\int_0^{2\pi} \cos(2\zeta)\, Q_{kin}(\zeta)d\zeta\right)\cos(2\phi_{2s}) \tag{S25}$$

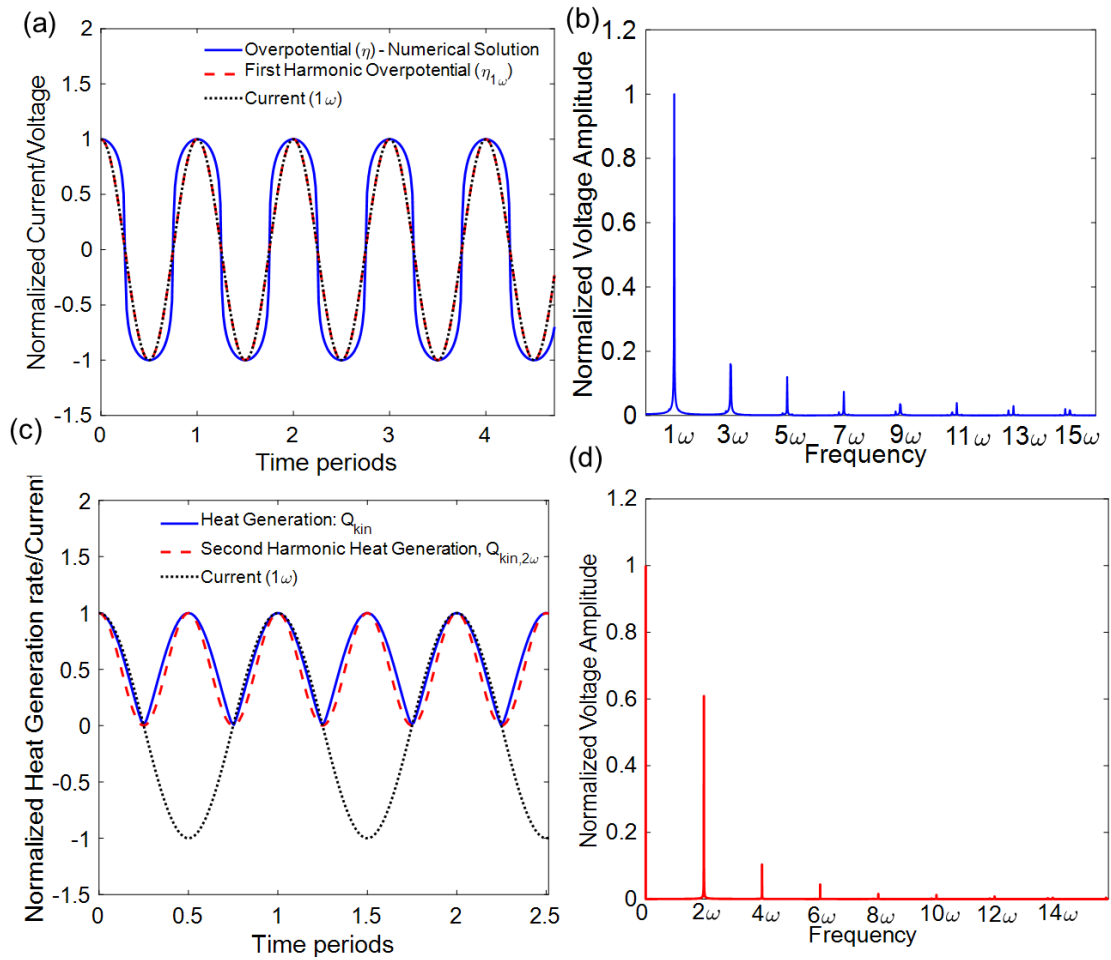

Figure S4. (a) Numerically solved overpotential (η) (blue solid line) plotted as a function of time for a sinusoidal current (black dotted line) along with the primary first-harmonic component of the overpotential (red-dashed line). The true overpotential (η) differs slightly from the first harmonic overpotential as it also has components in 3ω, 5ω and higher odd harmonics, which is illustrated in the Fourier transform of the overpotential shown in (b). The numerically solved heat generation rate due to the reaction kinetics ($Q_{kin}$) is shown in (c) with its Fourier transform shown in (d). Compared to the 1ω current (black dotted line in (c)), the heat generation rate has a DC offset and oscillated primarily at the second harmonic (2ω, red dashed line in (c)) but also has components in 4ω, 6ω and higher even harmonics.

*Heat generation associated with capacitance*

The current through the capacitive branch is $I_1 = I_{1,0} \cos(\omega t - \phi_{1c})$. The voltage drop across the capacitance is $\Delta V = \left(\frac{1}{\omega C_{dl}}\right) I_{1,0} \sin(\omega t - \phi_{1c})$. The product of the voltage across an ideal capacitor and the current is the energy stored/released by the ideal capacitor in the charge stored by the capacitor, and there is no heat generation/absorption associated with this process. Therefore,

$$Q_{dl,2\omega} = 0 \tag{S26}$$

The in-phase and the out-of-phase components are also:

$$Q_{dl,2\omega,IP} = 0 \tag{S27}$$

$$Q_{dl,2\omega,OP} = 0 \tag{S28}$$

The summary of all the heat generation terms at different harmonics are presented in Table S1.

Table S1. Summary of equations to calculate the heat generation rates pertaining to different electrochemical processes

| Harmonics | Process | Heat Generation Rate Magnitude | Eqn. |
|---|---|---|---|
| First Harmonic (1ω) | Entropic | $Q_{reversible,1\omega,IP} = -I_{2,0} \pi_{electrode} \cos(\phi_{2s})$ | S15 |
| | | $Q_{reversible,1\omega,OP} = I_{2,0} \pi_{electrode} \sin(\phi_{2s})$ | S16 |
| Second Harmonic (2ω) | Transport | $Q_{transport,electrolyte,2\omega,IP} = -\frac{I_0^2 R_t}{2}$ | S18 |
| | | $Q_{transport,SEI,2\omega,IP} = -\frac{I_{2,0}^2 R_{SEI}}{2} \sin(2\phi_{2s})$ | S20 |
| | | $Q_{transport,SEI,2\omega,OP} = -\frac{I_{2,0}^2 R_{SEI}}{2} \cos(2\phi_{2s})$ | S21 |
| | Charge-transfer kinetics | $Q_{kin,2\omega,IP} = \left(\frac{1}{\pi}\int_0^{2\pi} \cos(2\zeta) Q_{kin}(\zeta) d\zeta\right) \sin(2\phi_{2s})$ | S24 |
| | | $Q_{kin,2\omega,OP} = \left(\frac{1}{\pi}\int_0^{2\pi} \cos(2\zeta) Q_{kin}(\zeta) d\zeta\right) \cos(2\phi_{2s})$ | S25 |
| | Capacitance | $Q_{dl,2\omega,IP} = 0$ | S27 |
| | | $Q_{dl,2\omega,OP} = 0$ | S28 |

### 3. Frequency domain temperature rise at the sensor: Feldman's Algorithm

After calculating the heat generation rate caused by the various electrochemical processes, we need to be able to relate the effect of those heat generation rates to the frequency dependent surface temperature rise, which can be measured by a sensor. Albert Feldman [49] has provided a solution for the frequency dependent temperature rise at the surface of a stack with arbitrary number of layers due to a periodic planar heat source (units of W/m²) at an arbitrary location. Therefore, we will not re-discuss the solution here. However, if the heat generation is volumetrically distributed in a particular layer $J$, the surface temperature rise can be calculated by treating the solution provided by Feldman as a Green's function solution to the distributed heat generation rate. For a heat generation rate $Q$ (in Watts) distributed across a layer with thickness $L_j$ which is the $j^{th}$ layer in the stack, the surface temperature rise can be calculated as:

$$T_{surface} = \frac{Q}{A_s L_j} \int_0^{L_j} G(\zeta) \, d\zeta \tag{S29}$$

where, $G(\zeta)$ is the solution to the surface temperature due to unit strength (1W/m²) planar heat source at a location $\zeta$ in the $j^{th}$ layer of thickness $L_j$, i.e. $G(\zeta)$ has the units K/(W/m²)=m²K/W

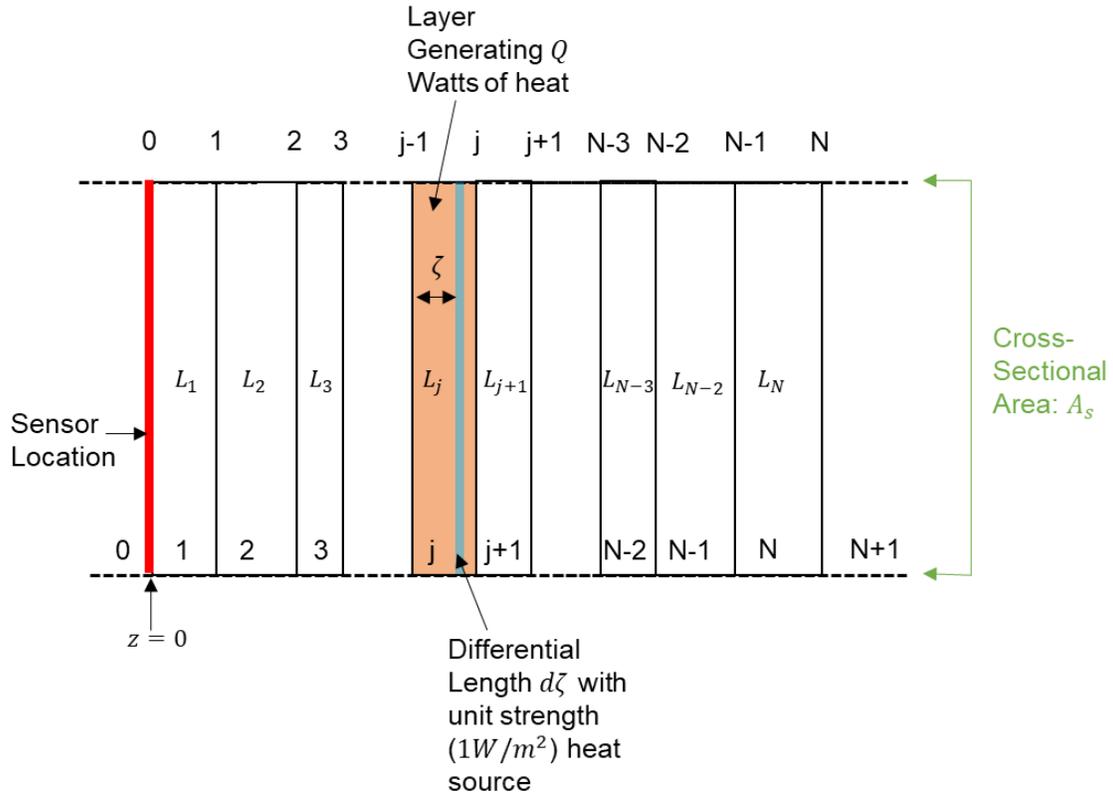

Figure S5. Schematic of the stack of N layers with a planar heat source at a location $\zeta$ in the $j^{th}$ layer of thickness $L_j$ as described in the solution to a periodic planar heat source by Feldman [49]. Feldman's solution ($G$) for the temperature rise at the sensor (at $z = 0$) due to a unit strength planar heat source can be generalized for a volumetric heat source of power $Q$ (watts) by integrating along the length of the layer as presented in Equation S29.

For an electrochemical cell, with the stack of positive current collector, cathode, separator/electrolyte, anode and the negative current collector, each layer and interfaces can have heat generation rates associated with the entropic, transport, capacitive and kinetic processes. The overall temperature response at the surface can be calculated as a sum of the individual temperature response of the individual processes. Additionally, in our analysis, the interfaces are treated as thin layers (arbitrarily chosen to be 1nm) with uniform heat generation. The choice of the interface thickness does not affect the temperature rise calculation as long as the interface heat capacity is low (chosen to be $1\ J/m^3K$) and the interface thermal resistance is small.

## 4. METS Fitting Algorithm

For any set of electrochemical properties, the heat generation rates at the layers and the interfaces can be calculated using equations presented in Table S1. If the thermal properties of each layer and the interfaces are known, then the surface temperature oscillations caused by these heat generation terms can be calculated using Feldman's algorithm, and the METS spectrum can be simulated for the entire range of frequency and current amplitudes used in the experiments. Then, from the best-fit between the experimentally measured temperature spectrum and the simulated spectrum, the electrochemical properties can be determined. The sensitivity and uniqueness of the fit are discussed in the results section, and the flowchart of the fitting algorithm for the METS measurements is presented in Figure S6.

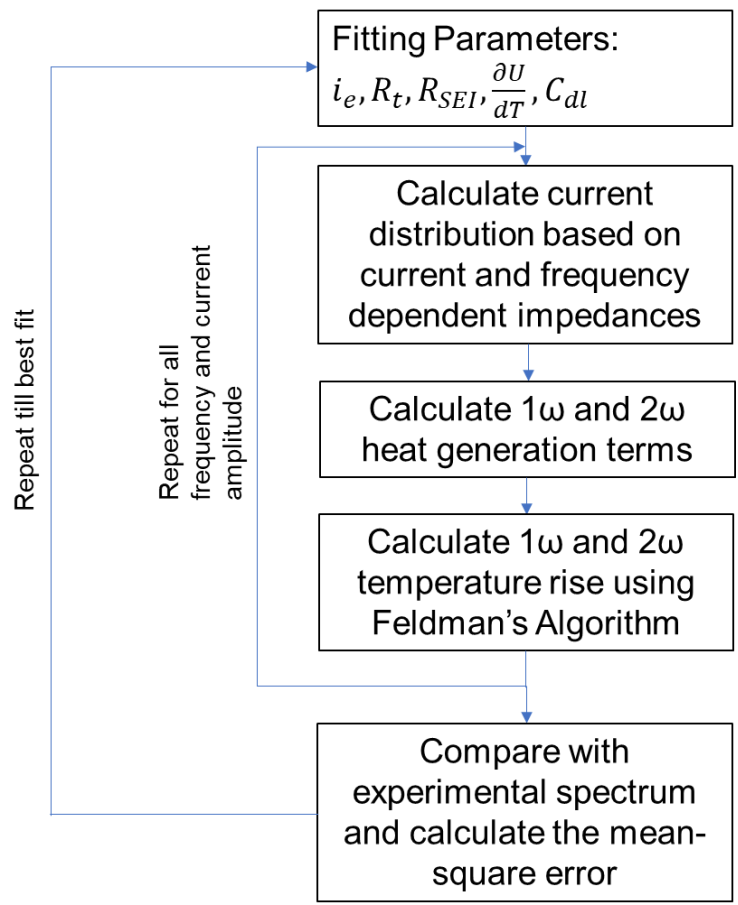

Figure S6. Flowchart of the METS fitting algorithm. Electrochemical properties of interest are determined from the best-fit between the simulated and experimentally measured METS spectrum.

## 5. Thermal treatment for interfacial heating in planar and porous electrodes

For the thermal analysis of planar electrodes, it was assumed that the interfacial heat generation occurred at the first 15nm of the electrodes. This choice was arbitrary and based on the assumption that most of the interfacial heat generation was in the SEI layer, whose length-scale is typically 10s of nm [56]. The choice of this length, however, does not affect the METS results significantly if the length is chosen to be thin enough (<1μm). Unlike in the planar electrode, the interface in porous electrode is distributed throughout the electrode. Therefore, we assume a uniform volumetric heating in the electrode due to the heat generation at the electrolyte-electrode interface. Due to non-uniform current distribution, it is possible that most of the heat generated is in the region closer to the separator than the region closer to the current collector, but in the frequency-range studied (0.1 Hz to 30 Hz), this spatial effect cannot be significantly differentiated within the measurement resolution.

## 6. Details of Experimental Methods

*Cell fabrication*

Symmetric cells (Figure S7 (b)) were made by sandwiching 1in × 1in lithium foil electrodes (MSE Supplies) between 10μm copper current collectors deposited on thermally conductive Kapton ® films with 25μm thick Celgard® 2400 separators in between. One of the dielectric films had the METS/3ω sensor deposited and wired. 2-3 mm thick Styrofoam sheet was attached on the sensor side of the cell to work as thermal insulation [47], [48] and a 2-3 mm thick Teflon plate was used on the other side to work as scaffolding. The cell was then sealed in a pouch cell configuration [47], [48] after adding the electrolyte (1M $LiPF_6$ in 1:1 EC:DEC, Sigma). NMC-Lithium full cell was made by using NMC-532 Cathode (MTI Corporation) with 60μm thick cathode and 15μm aluminum current collector. The sensor was placed on the lithium (anode) side and the stack configuration is presented in Figure S7 (d). Cell with one electrodeposited lithium electrode and one foil electrode cells were made by electrodepositing 15μm lithium on the sensor side current collector from the lithium foil used on the other side. The thickness of the copper current collector on both sides was 0.5μm and Styrofoam was used on both sides of the cell instead of Teflon on one side. NMC-graphite full cell with sensors on both anode and the cathode side were made with NMC 532 Cathode (MTI) and 60μm thick graphite anode with 11μm copper current collector (MTI Corporation). Teflon plate was used on both sides as scaffolding/insulation. Multilayer NMC-graphite cell was made by folding a 2 in × 1in NMC-graphite cell in half to make a 1 in × 1in Cathode-Separator-Anode-Anode-Separator-Cathode multi-stack with the sensor inserted in the middle. The NMC-Lithium and NMC-graphite cells were subject to three constant current formation cycles with cutoff-voltages between 4.3V and 3.0V. The charge/discharge current for NMC-lithium, single layer NMC-graphite and multilayer NMC-graphite cells were 1.5mA, 1.1 mA and 2.4 mA respectively and the capacities achieved after formation were 12.5mAh, 3.5mAh and 9.5mAh respectively.

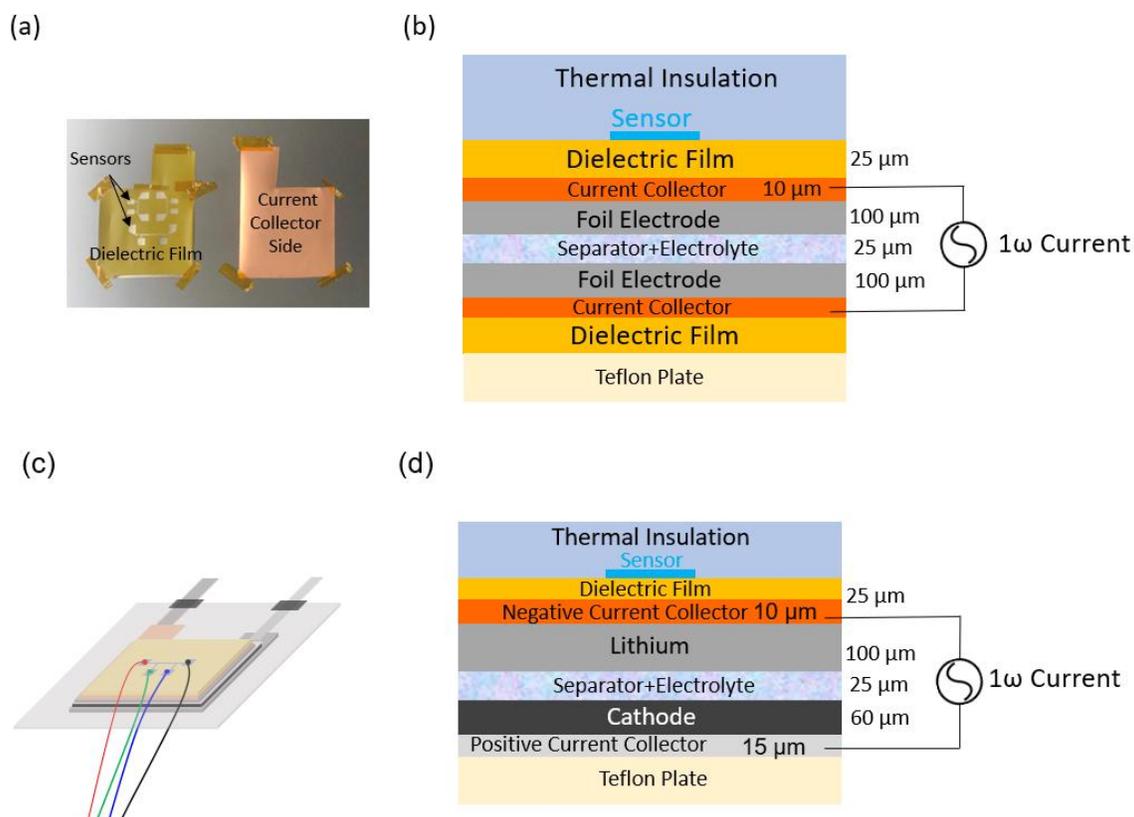

Figure S7. (a) METS sensors deposited on one side of a dielectric film (left) and a copper film acting as the current collector deposited on the other side of the dielectric film (right), (b) schematic of the model cell with symmetric foil electrodes and a sensor on one side, (c) pouch cell assembly of the cell stack with the sensor and (d) ) schematic of a cell with NMC cathode and lithium foil anode with the sensor on the lithium side.

Table S2. Summary of different cells examined.

| Cell | Cell Type | Electrode close to the sensor | Thickness (μm) | Electrode away from the sensor | Thickness (μm) |
|---|---|---|---|---|---|
| 1 | Symmetric lithium | Lithium foil | 100 | Lithium foil | 100 |
| 2 | NMC-Lithium | Lithium Foil | 60 | NMC 532 | 60 |
| 3 | Symmetric with electrodeposited and foil lithium | Electrodeposited lithium | 15 | Lithium foil | 85 |
| 4 | NMC-Graphite (Single Layer) | Graphite | 50 | NMC 532 | 60 |
| 5 | NMC-Graphite (Multilayer) | Graphite | 50 | NMC 532 | 60 |

*High-precision thermometry instrumentation*

The temperature oscillations associated with METS signal are of the order of mK. Therefore, it is necessary to implement instrumentation that minimizes external noise and allows the isolation of the signal at a particular frequency. Frequency dependent (lock-in based) temperature measurements of the order of a few µK have been conducted before [57], [58], [59]. In these measurements, either a full-bridge or a half-bridge circuit to cancel the dominant off-frequency components and noise have been implemented. In our case, we use a half-bridge circuit with a matching resistor to cancel the dominant DC voltage across the sensor. In our measurements, the signal is generated by passing an alternating current at a frequency ω through the cell, which causes temperature oscillations at the sensor at frequencies 1ω and 2ω. Due to linear temperature dependence of the resistance, the temperature oscillations at 1ω and 2ω cause resistance oscillations at the sensor at the frequencies 1ω and 2ω, i.e. $R_{1\omega} = \left(\frac{dR}{dT}\right)T_{1\omega}$ and $R_{2\omega} = \left(\frac{dR}{dT}\right)T_{2\omega}$, where $\left(\frac{dR}{dT}\right)$ is the linear temperature coefficient of resistance of the sensor. The sensor, which has a constant DC current passing through it, experiences voltage oscillations at 1ω and 2ω from the relation $V_{1\omega} = I_{DC}R_{1\omega}$ and $V_{2\omega} = I_{DC}R_{2\omega}$. This oscillating voltage can be measured using a lock-in amplifier after the predominant DC voltage is cancelled using a matching resistor.

The schematic of the signal generation and instrumentation for the measurement is shown in Figure S8. In our measurements, we use two Keithley 6221 current sources, one as the AC source for the cell and one as a DC source for the sensor. The frequency of the AC source is referenced to a SR830 lock-in amplifier, which measures the voltage oscillations across the sensor. Typically, a lock-in amplifier multiplies the input signal with the chosen harmonic of the reference signal to reconstruct the amplitude of the input signal at the chosen harmonic as a DC signal, which is then passed through a low-pass filter to isolate and extract as a demodulated signal. The demodulated signal which is an output of the lock-in amplifier contains both the magnitude and the phase difference between the input signal and the reference frequency, which the lock-in can output as the in-phase and out-of-phase voltages (and consequently the temperature using temperature coefficient of resistance) used in the analysis. We direct the readers to the manual of the SR830 lock-in amplifier [60] used in this work to get an overall understanding of the principles behind signal demodulation in a lock-in amplifier.

In all of our experiments, the typical noise in the voltage is within 100-200 nV. Considering the typical current through the sensor to be 10 mA and typical temperature coefficient of resistance of the sensor to be 0.15 Ω/K (varies slightly for each sensor), this voltage noise translates to a noise in temperature measurement of ~65-150µK. In future experiments, we believe that this noise can be minimized further by implementing a full-bridge cancellation circuit [59] and by using co-axial cables or twisted pairs for signal transmission.

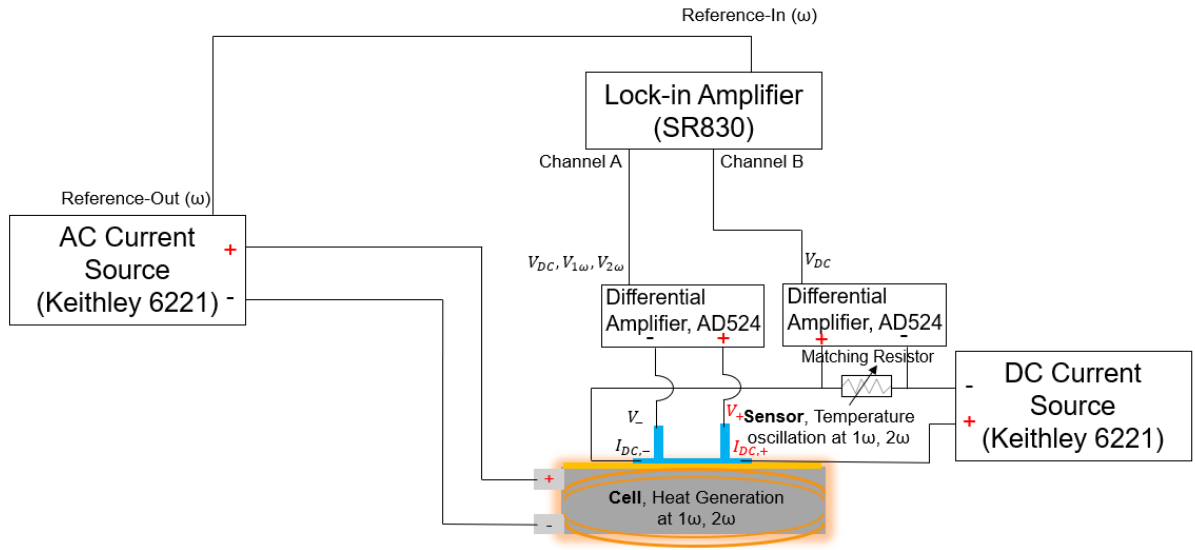

Figure S8. Schematic of the signal generation and measurement. The heat is generated by passing AC current through the cell, which causes temperature oscillations at the sensor. The frequency of the AC source is referenced to the lock-in amplifier which measures the voltage oscillation across the sensor through Channel A and the DC offset voltage simulated by a matching variable resistor through Channel B. The DC offset is subtracted out in the final measurement by taking the measurement in the mode A-B. Unit gain differential amplifiers (AD524) are used to ensure the grounds of signal going to A and B are referenced to the same voltage.

*3ω measurements and thermal property characterization*

The same sensor and the setup used for METS signal acquisition can be used for traditional 3ω measurements. However, instead of passing the alternating current through the cell, the alternating current is passed through the sensor itself, creating a 2ω temperature fluctuation and a 3ω voltage fluctuation at the sensor, which can be used to measure the thermal transport properties of the layers and interfaces [61], [62]. After the cell is assembled, before the METS experiment, we perform a 3ω experiment to determine the effective thermal resistance of the interfaces, which is then used along with the thermal properties of the layers to calculate the frequency dependent temperature rise using Feldman's algorithm. To minimize the uncertainty in the 3ω measurement of the interface resistance, the thermal properties of each layer are predetermined either from individual 3ω measurements for thermal conductivity and differential scanning calorimetry (DSC) and density measurement for heat capacity or taken from literature. In the case of symmetric cells and the NMC-lithium cells, we do not have specific sensitivity to individual lithium-separator and cathode-separator interfaces. So, we assume both the electrode-separator interfaces have the same thermal resistance and fit a single value of resistance to match the 3ω spectrum. However, in the case of one electrodeposited and one foil electrode, we observe that the lithium-separator interface resistance is much smaller at the deposited electrode-separator interface than at the foil-electrode separator interface and therefore can individually fit the interface resistances to match the 3ω spectrum. We believe the interface resistance is higher for the foil-lithium separator interface because of the pre-existing macroscopic non-homogeneities (roughness) at the foil lithium surface.

The thermal properties of the layers and interfaces used in METS analysis are summarized in Table S4.

## 7. 3ω measurements for thermal interface resistance

The 3ω fitting and summary of results are presented in Figure S9 and Table S3 respectively.

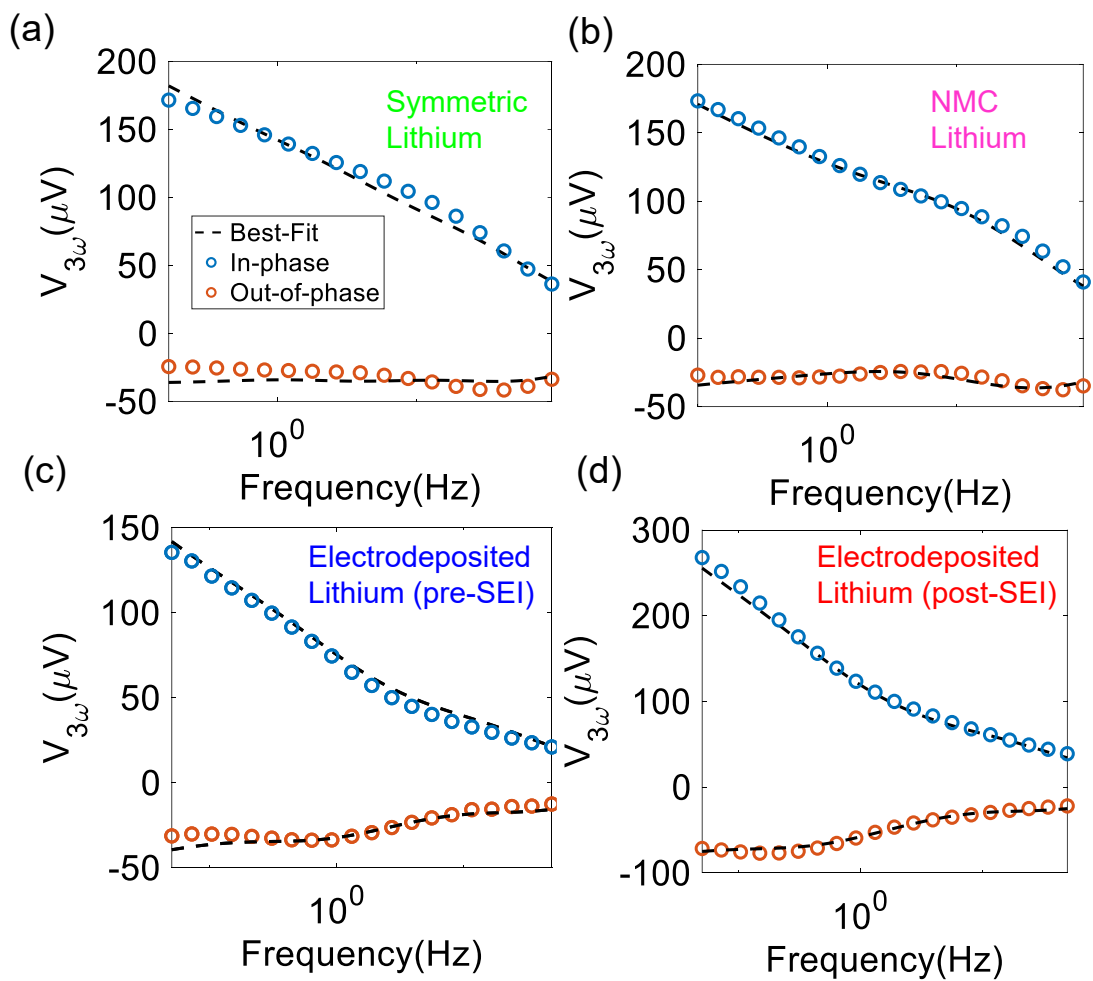

Figure S9. The in-phase (blue circles) and out-of-phase (orange circles) 3ω measurements along with the best-fit spectrum (black dashes) to the 3ω measurements for determining thermal interface resistances in (a) symmetric lithium cell, (b) NMC-lithium cell, (c) electrodeposited lithium cell before SEI growth and (d) electrodeposited lithium cell after SEI growth. The summary of the best-fit interface resistances are presented in Table S3.

Table S3. Summary of the best-fit thermal interface resistances

| Cell | Lithium-Copper Interface Resistance (sensor side) | Electrode (lithium)-separator interface resistance (sensor side) | Electrode - separator interface resistance (non-sensor side) | Electrode-current collector interface resistance (non-sensor side) |
|---|---|---|---|---|
| Lithium Symmetric | 5 cm$^2$K/W | 5 cm$^2$K/W | 5 cm$^2$K/W | 5 cm$^2$K/W |
| NMC-Lithium | 1 cm$^2$K/W | 3.7 cm$^2$K/W | 3.7 cm$^2$K/W | Negligible |
| Electrodeposited lithium-Foil lithium (pre-SEI growth) | Negligible | 0.02 cm$^2$K/W | 12.5 cm$^2$K/W | 12.5 cm$^2$K/W |
| Electrodeposited lithium-Foil lithium (post-SEI growth) | Negligible | 0.02 cm$^2$K/W | 25 cm$^2$K/W | 25 cm$^2$K/W |

## 8. Summary of thermal properties

Table S4. Summary of thermal properties of each layer and interfaces

| Layer/Interface | Thermal conductivity (W/mK) | Volumetric heat capacity (MJ/m$^3$K) | Reference |
|---|---|---|---|
| Styrofoam (Insulation) | 0.1** | 0.175* | |
| Thermally Conductive Kapton | 0.48* | 1.84* | |
| Copper film | 401 | 3.44 | [22] |
| Copper-lithium interface | 3ω best-fit | - | |
| Lithium metal | 85 | 1.913 | [63] |
| NMC-cathode-separator interface | 3ω best-fit | - | |
| Lithium-separator interface | 3ω best-fit | - | |
| Separator + electrolyte | 0.3 | 2.180 | [48] |

*measured, **estimated

# 9. Phase relationship between current, heat generation rate and temperature rise in METS experiments

The temperature rise at the sensor measured by the lock-in amplifier is resolved into in-phase and out-of-phase components with reference to the current passed through the cell. In order to develop an intuitive understanding of the phase relationship between the applied alternating current and the measured temperature oscillations, it is important to understand the phase relationship between the heat generation rate and the temperature rise at the sensor.

A battery with a METS sensor can be simplified as a layered structure with a sensor at the one end of the stack and heat generating layers within the stack as shown in Figure S10. If sinusoidal heat is generated at a layer adjacent to the sensor so that there is no thermal conduction lag between the sensor and the heat generating layer, as represented by the blue heat generating layer and the black sensor in Figure S10, the temperature rise at the sensor can be related to the heat generation rate ($Q$) through an effective heat capacity, i.e. $\rho C_p \left(\frac{dT}{dt}\right) = Q$. If the heat generation rate is sinusoidal i.e. $Q \sim \sin(\omega t)$, then the temperature rise is of the form $T \sim \cos(\omega t)$, i.e. the temperature rise lags the heat generation rate by exactly 90 degrees. In other words, for sinusoidal heating, if there is no thermal conduction lag between the heat source and the sensor, the temperature rise lags the heat generation rate by 90 degrees because of thermal capacitance. This is illustrated in the plot in Figure S10, where the temperature rise at the sensor (shown in blue solid line) lags the heat generation at the layer adjacent to the sensor (shown in black dotted line) by 90 degrees. If the heat generation sources are further away from the sensor, such as in layers 3 and 5 shown in green and red color in the schematic in Figure S10, there is an additional phase lag between the heat generation rate (black dotted line) and the temperature rise at the sensor represented by the green dashed line and the red dashed line for heating in layer 3 and layer 5 respectively.

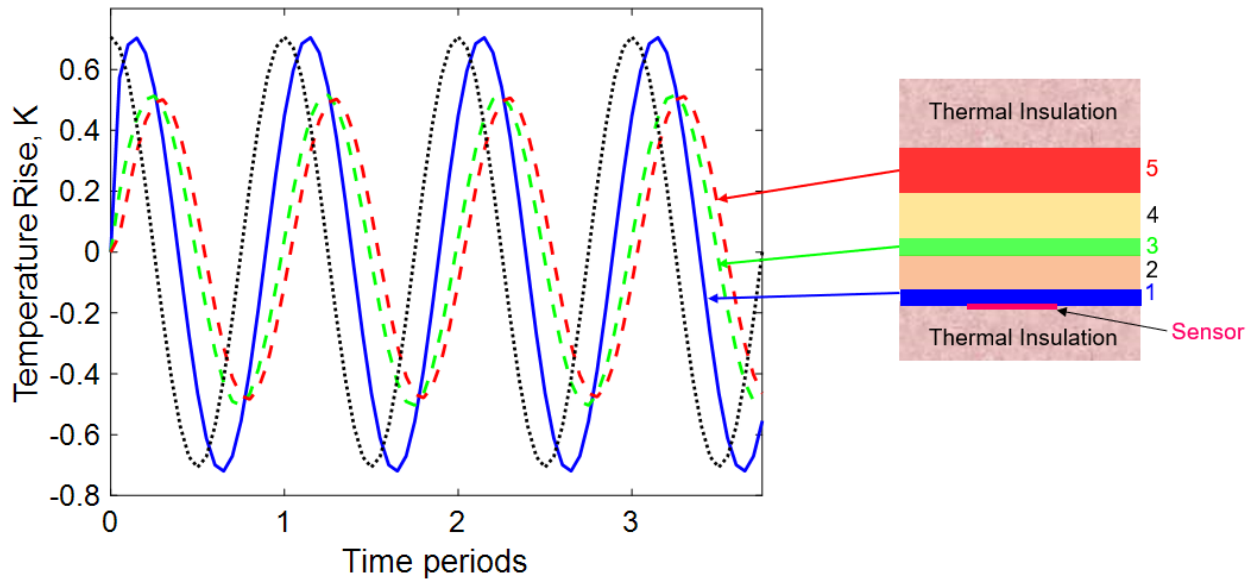

Figure S10. Left: Oscillating temperature rise at the sensor plotted as a function of time for time-dependent sinusoidal heating (black-dotted line) at layer 1 (blue solid line), layer 3 (green dashed line) and layer 5 (red dashed line) for a layered structure with a sensor (pink) on the bottom and thermal insulation around it, shown on the right. When the heating location is adjacent to the sensor, i.e. at layer 1, there is no thermal conduction lag between the heat source and the sensor, and the temperature rise lags the heat generation by exactly 90 degrees because of the thermal capacitance of the layers, as illustrated by the phase difference between the black dotted line (heat generation) and the blue solid line (temperature rise). For heat sources away from the sensor, the time lag due to thermal conduction creates an additional phase lag in the temperature rise at the sensor with respect to the phase of the heating, which is evident from the additional phase lag seen in the green-dashed and red-dashed lines corresponding to the respective heating in layer 3 and layer 5 in the structure shown on the right.

For 1ω heat due to entropy change at the electrode-electrolyte interface, the heat generation is proportional to the current and is therefore at the same phase of the applied current (except for a small phase difference ($\phi_{2c}$) due to the current distribution). Thus, the temperature rise, being 90 degrees out-of-phase from the heat generation rate, is primarily observed as out-of-phase signal with respect to the applied current, with any in-phase component arising from the thermal lag between the heat generating interface and the sensor. This is observed in Figure 2 (a) and 2 (b). Similarly, for the 2ω temperature, shown in Figure 4 (a)-(f) and Figure 5 (a)-(b), the heat generation is proportional to the square of the current. Therefore, for a sinusoidal 1ω current, the 2ω heat is of the form $\cos(2\omega t)$, and the corresponding temperature rise is of the form $\sin(2\omega t)$. Therefore, the temperature rise is primarily in-phase with the current and the out-of-phase component can be attributed to the thermal conduction lag between the heat generating layer/interface and the sensor.

## 10. Verification of the thermal analysis and Feldman's algorithm

Before proceeding into validating the electrochemical and thermal aspects of METS and studying electrochemical systems using METS, we first verified that the thermal model of relating frequency modulated heat generation with the surface temperature rise (i.e. Feldman's method) and data acquisition (i.e. lock-in based resistance thermometry) is correct. To do so, we deposited a METS sensor on one side of a dielectric film and a serpentine resistive heater on the other side of the dielectric film, shown in Figure S11 (b). The thermal conductivity of the dielectric film was determined using the 3ω method [48], [64] (Figure S11c) by using the METS sensor as a 3ω sensor. An alternating current of a constant amplitude $I_0$ was passed through the resistive serpentine heater to cause a 2ω heat generation (because of $I^2R$ heating) at the heater and a corresponding 2ω temperature oscillation at the sensor. The schematic of the experimental stack is shown in Figure S11 (d). This temperature oscillation was measured via resistance thermometry by measuring the 2ω voltage oscillation using the lock-in amplifier. After obtaining the experimental frequency spectrum of the 2ω temperature, we used Feldman's method to simulate the 2ω temperature spectrum for a chosen value of the resistance $R$ and a known value of the current amplitude $I_0$. The best-fit between the simulated 2ω spectrum and the measured 2ω spectrum was obtained when the value of the resistance $R$ was 910 Ω and is shown in Figure S11 (a). From an independent 4-point electrical resistance measurement, we measured the resistance of the heater to be 863.8 Ω, which is within 5% of the value determined from the best-fit. Since the directly measured resistance value was within 5% of the value estimated from the thermal analysis, we were able to verify the general accuracy of the thermal analysis (Feldman's algorithm) and accuracy of the experimental instrumentation.

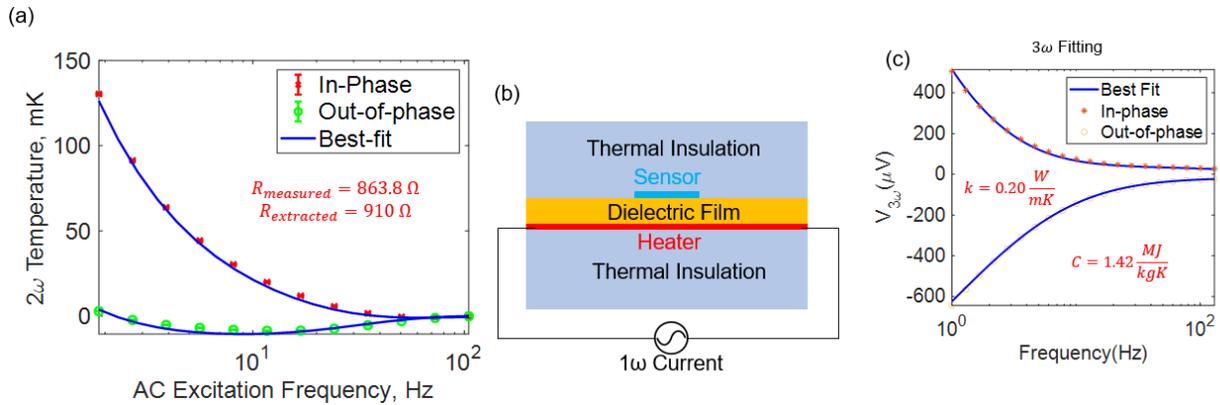

Figure S11. (a) Best-fit METS spectrum between the experimental (red crosses: in-phase and green circles: out-of-phase) temperature measurements and the simulated (blue solid lines) temperature spectrum for modulated 2ω heating using a resistive heater, (b) ) schematic of the resistance heater-METS sensor setup with serpentine heater (bottom) and a METS/3ω sensor (top) and (c) 3ω best-fit to determine the thermal conductivity of the dielectric (Kapton ®) film and

## 11. METS fits for the electrodeposited cells

Due to space constraints, the 2ω METS fitting for the cells with one electrodeposited and one foil lithium electrodes for different current amplitudes could not be presented in the main text and are therefore presented below.

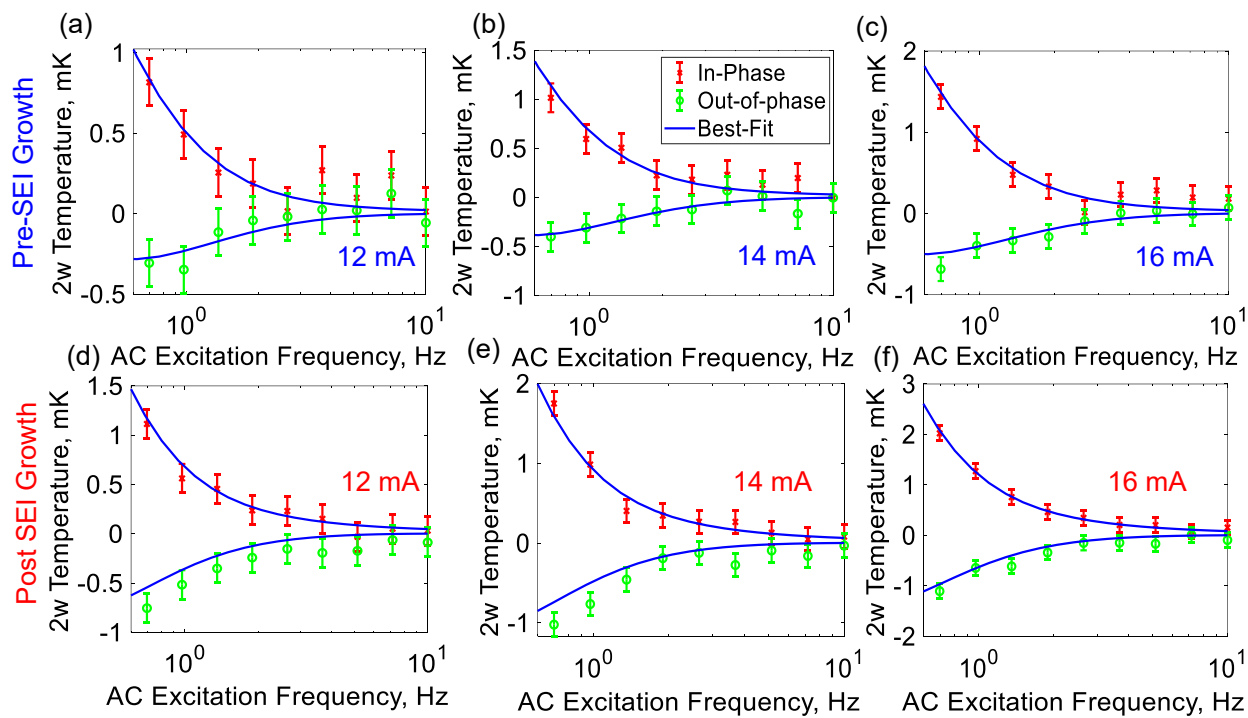

Figure S12. The 2ω temperature spectrum showing in-phase (red cross) and out-of-phase (green-circles) temperature rise as a function of the frequency of the current passed through the cell for current amplitudes of 12mA, 14mA and 16mA for the cell with one electrodeposited lithium electrode and one foil lithium electrode before SEI growth (a-c, top) and after SEI growth (d-f, bottom). The negative out-of-phase signal implies that the majority of the signal is generated at the interface away from the sensor, indicating that the resistance at the interface closer to the sensor is much smaller than the resistance at the interface away from the sensor. Additionally, the magnitude of the temperature for the same current magnitudes is seen to increase in (d)-(f) (post-SEI growth) when compared with (a)-(c) (pre-SEI growth) indicating the increase in Ohmic heat due to SEI growth.

## 12. METS on multi-cell battery

The measurements presented in this work are performed on a single-stack cell with the sensor on one side of the stack. The thermal analysis for the temperature measurement at the sensor placed on the end of the stack is based on the temperature solution provided by Feldman. However, the method of Feldman is not restricted to the sensor location at the boundary of the stack and can be extended to an arbitrary sensor position. We have presented the theoretical formulation for the temperature rise at a sensor in an arbitrary location in the stack in our other work [65]. The use of this formulation allows METS measurement to be done in a multi-cell battery with the sensor placed arbitrarily within the stack. The sensitivity of the measurement will be to the layers close to the stack. This is both advantageous and disadvantageous. The advantage is that the measurement allows measurement of locally non-homogeneous phenomena occurring near the sensor. The disadvantage is that the locally sensed information cannot be extended to layers further away from the sensor, requiring the use of multiple sensors for additional information.

To verify the use of METS in a multi-cell battery, we constructed a two-cell battery with NMC523 cathode and graphite anode. The stacking was cathode-separator-anode-anode-separator-cathode, with the METS sensor inserted in between the two anode layers. To simplify the METS and EIS results for verification purpose, we assume the two cathode layers behave similarly with the same resistance drop and state of charge. Similarly, we assume that the two anode layers also behave similarly. The cells were charged to a SOC of 0.5 with an open circuit voltage (OCV) of 3.5 V, similar to the single layer NMC-Graphite cell presented in the paper. Figure S13 shows the measured and best-fit 1ω (a) and 2ω (b) temperatures along with the corresponding 3ω thermal measurement and the EIS measurements (d) for the cell. The best-fit for the 1ω measurement is obtained for $\frac{dU}{dT} = 1.1 \pm 0.11$ mV/K for the anode and for $\frac{dU}{dT} = 1.3 \pm 0.23$ mV/K for the cathode, which are within 5% of the values measured for anode and cathode on the single layer NMC-graphite cell. In the EIS (S13 d), two prominent semi-circles corresponding to resistances 1.1 Ω and 6.4Ω can be seen, corresponding to the area specific resistance (per inch$^2$ electrode) to be 2.2 Ω and 12.8 Ω. The best-fit 2ω measurement is obtained for cathode resistance of $10 \pm 2.8$ Ω and anode resistance of $0.5 \pm 0.14$ Ω, assuming a small charge-transfer resistance for both (0.2 Ω), indicating that the smaller semi-circle observed in the EIS corresponds to the transport resistance at the anode, while the larger semi-circle corresponds to that at the cathode.

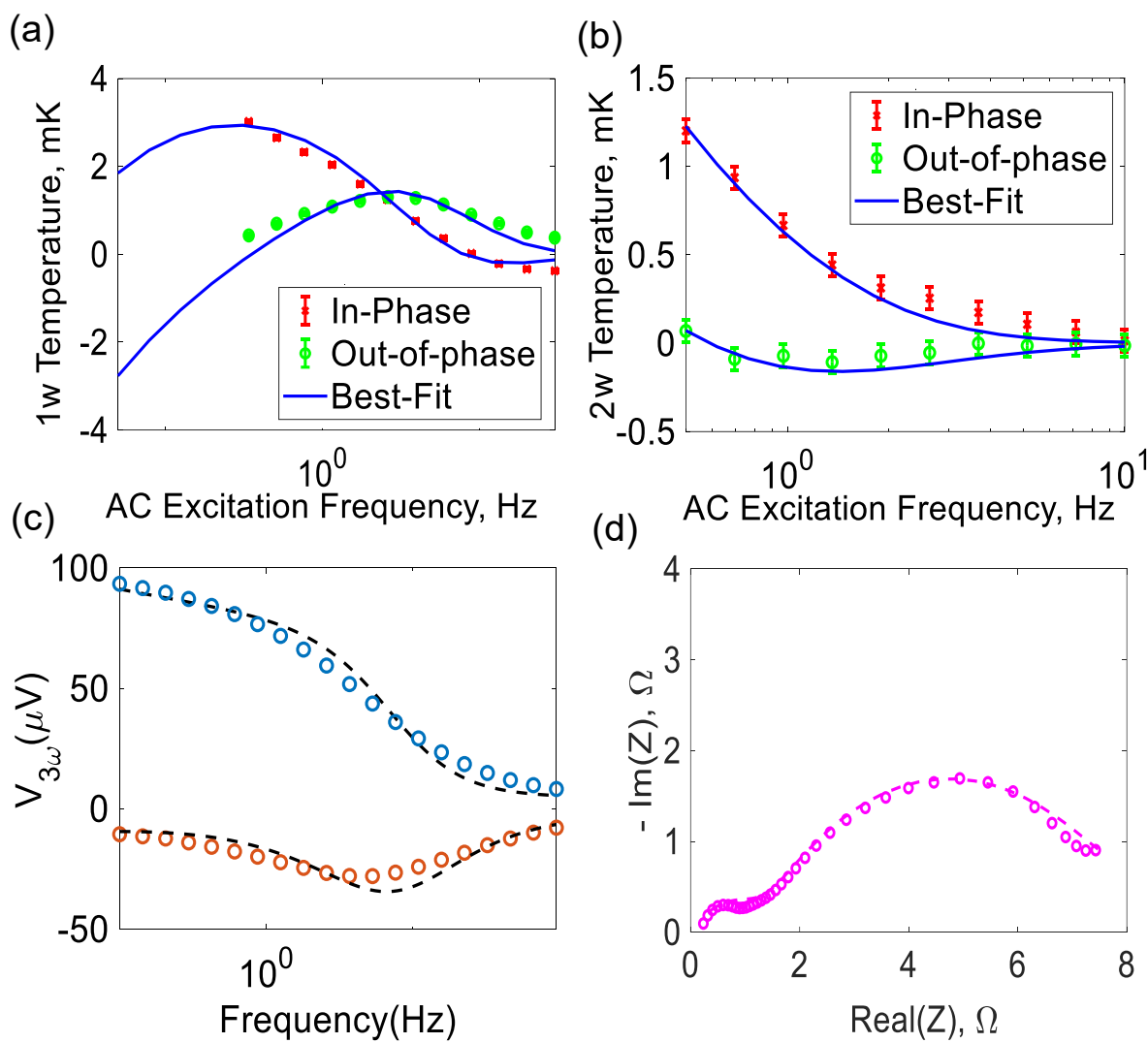

Figure S13. For a multi-cell stack with the METS sensor placed between two anode layers, the measured 1ω (a) and 2ω (b) plots with in-phase (red cross) and out-of-phase (green-circles) temperature rise and the best-fit lines (in blue). The 3ω measurement for thermal resistance determination is presented in (c), and the EIS measurement on the cell is presented in (d). The 1ω behavior is consistent with the anode side measurement performed on the single layer NMC-graphite cell. The 2ω best fit is obtained for the anode resistance of $0.5 \pm 0.14$ Ω and cathode resistance of $10 \pm 2.8$ Ω, which is close to the total resistance of 14.6Ω measured from EIS (d).

## 13. Effect of surface impurities on interface resistance of symmetric electrodeposited cell and foil cell

To examine the possibility of the surface impurities causing a higher interface impedance in the foil electrode compared to the electrodeposited electrode, as hypothesized in the interpretation of the measurements on cells with one electrodeposited and one foil electrode, we prepared two cells with electrodeposited electrodes on one side and foil electrodes on the other. Then, we isolated the foil electrodes and the electrodeposited electrodes from both cells to prepare two new cells, one with symmetric foil electrodes and one with symmetric electrodeposited electrodes. The EIS bode plot for the two cells are presented in Figure S13. As seen, the impedance of the cell with symmetric foil electrodes (red-circles) is much higher than that of the cell with symmetric electrodeposited electrodes (blue circles). This strengthens our hypothesis that the foil electrode contains surface impurities, which are not present in the electrodeposited electrode, which is generated by depositing pure lithium.

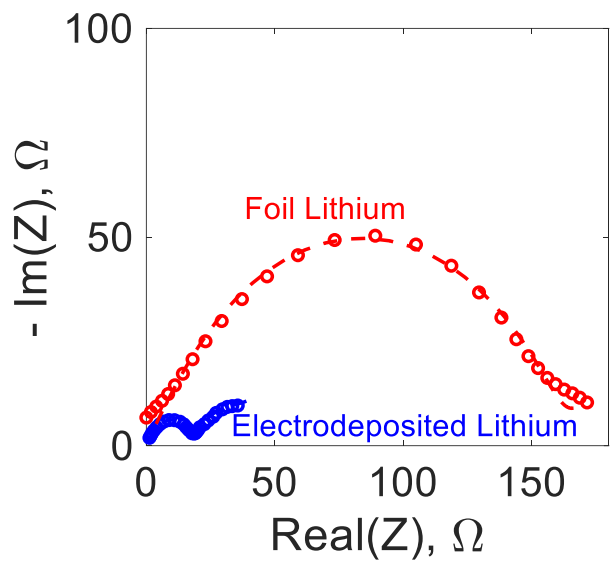

Figure S14. EIS spectrum for symmetric cells with two foil electrodes (red circles) and two electrodeposited electrodes (blue circles). The impedance of the cell with symmetric foil electrodes is much higher than that of the cell with symmetric electrodeposited confirming that the high impedance is caused by the surface impurities in foil electrodes as they are not present in the electrodeposited electrode generated by depositing pure lithium.

## 14. Calculation of measurement uncertainty

The measurement uncertainty ($U_p$) in each parameter ($p$) measured from METS by fitting the 1ω or 2ω spectrum can be calculated as the sum of uncertainties in all input parameters weighted by their sensitivity in the 1ω or 2ω spectrum [66], i.e.

$$U_p = \frac{\sum_{i=1}^{n, i \neq p} S_i U_i}{S_p} \tag{S30}$$

The uncertainty in the input parameters can be in their thermal properties or the electrochemical properties. The electrochemical properties and their respective uncertainties used in the calculations are presented in Table S5 and the thermal properties and their respective uncertainties are presented in Table S6.

Table S5. Uncertainties in electrochemical properties

| Property | Value | Uncertainty |
|---|---|---|
| Interface double layer capacitance | EIS best-fit | 10% [46] |
| Electrolyte conductivity | 7.2 mS/cm [67] to 2.7 mS/cm [68], Avg. 4.95 mS/cm | 45% [67], [68] |
| Lithium metal electrical conductivity | 1.08×10$^7$ S/m [69] | <1% ** |
| Aluminum electrical conductivity | 3.5×10$^7$ S/m [69] | <1% ** |
| Copper electrical conductivity | 5.96×10$^7$ S/m [69] | <1% ** |

**estimated

Table S6. Uncertainties in thermal properties

| Layer/Interface | k (W/mK) | Δk/k | C (MJ/m$^3$K) | ΔC/C | L (μm) | ΔL/L |
|---|---|---|---|---|---|---|
| Styrofoam (Insulation) | 0.1** | 20%** | 0.175** | 8% ** | 2000* | 50%* |
| Platinum (Sensor) | 169 [22] | 10%** | 2.85 [22] | 10%** | 100** | 10%** |
| Thermally Conductive Kapton | 0.484* | 1%** | 1.84* | 2.5%* | 25* | 1%* |
| Copper film | 401 [22] | 5% [22] | 3.44 [22] | 5% [22] | 5* | 10%* |
| Copper-lithium interface | 3ω best-fit | 10%** | N/A | N/A | N/A | N/A |
| Lithium metal | 85 [63] | 5%** | 1.913 [23] | 5% ** | 100 [23] | 2% [23] |
| Aluminum Current Collector | 237 [22] | 5% [22] | 2420 [22] | 5% [22] | 15* | 1%* |
| NMC Cathode | 1 [22] | 20% [48] | 3510 [22] | 10% [22] | 60* | 5% [22] |
| Graphite Anode | 1.1 [22] | 20% [22] | 1837 [22] | 10% [22] | 50* | 5% [22] |
| cathode-separator or anode-separator interface | 3ω best-fit | 10%** | N/A | N/A | N/A | N/A |
| Lithium-separator interface | 3ω best-fit | 10% ** | N/A | N/A | N/A | N/A |
| Separator + electrolyte | 0.3 [48] | 19% [48] | 2.180 [48] | 6% [48] | 25* | 3.8% [22] |

*measured in this work, **estimated

The uncertainty in 3ω fitting for the thermal interface resistance is estimated from 80% confidence interval in the interface resistance measurement presented in Lubner et al. [48]. As presented in Figure 2 (c)-(d) and Figure 5 (a)-(b) in the main text, the measurement sensitivity for each parameter is a function of the frequency. Therefore, the measurement uncertainty from equation S30 is also a function of frequency. Since the best-fit for a parameter is obtained at the frequency at which the measurement is the most sensitive for the parameter, the corresponding uncertainty reported in this work is also reported for the frequency at which the measurement sensitivity is the maximum. Additionally, the error in the reported resistances from EIS are assumed to be 10% for all measurements [46].